\documentclass{SCGEcontent}
\usepackage{titletoc}
\begin{document}

\begin{picture}(0,0){\rm
\put(0,-20){\makebox[160truemm][l]{\bf {\sanhao\raisebox{2pt}{.}}
Invited Review  {\sanhao\raisebox{1.5pt}{.}}}}}
\put(0,-34){\jiuwuhao {\textcolor[rgb]{0.5,0.5,0.5}{\sf Special Topic: the Next Detectors for Gravitational Wave Astronomy
}}}
\end{picture}

\def\bm{\boldsymbol}

\def\dl{\displaystyle}
\def\du{\end{document}}
\def\d{{\rm d}}
\def\e{{\rm e}}
\def\r{{\bm r}}
\def\P{{\bm P}}
\def\A{{\bm A}}
\def\k{{\bm k}}
\def\Q{{\bm Q}}
\def\pi{{\uppi}}
\def\cp#1{\mathbf{#1}}

\Year{2015} %
\Month{December} %
\Vol{X} 
\No{X} 
\BeginPage{1} 
\EndPage{11} 
\AuthorMark{{\rm Blair D}, et al.}  
\AuthorMarkCite{{\rm Zhang Y, Liu Y X, Hou Z F}, et al.} 
\DOI{10.1007/s11433-015-5747-7} 
\ArtNo{X}

\title{The next detectors for gravitational wave astronomy\footnotemark[2]\footnotetext[2]{Sect. 1 is contributed by BLAIR David, JU Li, ZHAO ChunNong, WEN LinQing, MIAO HaiXing,
CAI RongGen, GAO JiangRui, LIN XueChun, LIU Dong, WU Ling-An, ZHU ZongHong
(corresponding author, BLAIR David, email: david.blair@uwa.edu.au); sect. 2 is
contributed by HAMMOND Giles (email: giles.hammond@glasgow.ac.uk); sect. 3 is contributed by PAIK Ho Jung (email: hpaik@umd.edu); sect. 4 is contributed by FAFONE Viviana, ROCCHI Alessio (FAFONE Viviana, email:
viviana.fafone@roma2.infn.it; ROCCHI Alessio, email: alessio.rocchi@roma2.infn.it); sect. 5 is contributed by BLAIR Carl (email: carl.blair@uwa.edu.au); sect. 6 is
contributed by ZHAO ChunNong, MA YiQiu, QIN JiaYi, PAGE Michael, JU Li, BLAIR David (corresponding author, ZHAO ChunNong, email: chunnong.zhao@uwa.edu.au); sect. 7 is contributed by MIAO HaiXing (email: haixing@star.sr.bham.ac.uk)}}

\author[1]{BLAIR David}{}
\author[1]{JU Li}{}
\author[1]{ZHAO ChunNong}{}
\author[1]{WEN LinQing}{}
\author[2]{MIAO HaiXing}{}

\author[3]{CAI RongGen}{}
\author[4]{\vspace{1.3mm}\\GAO JiangRui}{}
\author[5]{LIN XueChun}{}
\author[6]{LIU Dong}{}
\author[7]{WU Ling-An}{}
\author[8]{ZHU ZongHong}{}

\author[9]{\vspace{1.3mm}\\HAMMOND Giles}{}
\author[10]{PAIK Ho Jung}{}
\author[11,12]{FAFONE Viviana}{}
\author[12]{ROCCHI Alessio}{}
\author[1]{\vspace{1.3mm}\\BLAIR Carl}{}
\author[1]{MA YiQiu}{}
\author[1]{QIN JiaYi}{}
\author[1]{PAGE Michael}{}

\address[{\rm1}]{School of Physics, The University of Western Australia, Crawley WA 6009, Australia;}
\address[{\rm2}]{School of Physics and Astronomy, University of Birmingham, Birmingham B15 2TT, UK;}

\address[{\rm3}]{State Key Laboratory of Theoretical Physics, Institute of Theoretical Physics, Chinese Academy of Sciences, Beijing 100190, China;}
\address[{\rm4}]{The School of Physics \& Electronic Engineering, Shanxi University, Taiyuan 030006, China;}
\address[{\rm5}]{Laboratory of All-Solid-State Light Sources, Institute of Semiconductors, Chinese Academy of Science, Beijing 100083, China;}
\address[{\rm6}]{State Key Laboratroy of Modern Optical Instrumentation, Department of Optical Engineering, Zhejiang University, Hangzhou 310027, China;}
\address[{\rm7}]{Laboratory of Optical Physics, Institute of Physics, Chinese Academy of Sciences, Beijing 100190, China;}
\address[{\rm8}]{Gravitational Wave and Cosmology Laboratory, Department of Astronomy, Beijing Normal University, Beijing 100875, China;}

\address[{\rm9}]{School of Physics and Astronomy, SUPA University of Glasgow, Glasgow, UK;}
\address[{\rm10}]{~~Department of Physics, University of Maryland, College Park, MD 20742, USA;}
\address[{\rm11}]{~~University of Rome Tor Vergata, Rome 00133, Italy;}
\address[{\rm12}]{~~INFN Roma Tor Vergata, Rome 00133, Italy}

\maketitle \vspace{-3.5mm}{\footnotesize\begin{center} Received September 24, 2015; accepted September 28, 2015
\end{center}}\vspace*{-5mm}

\begin{center}
\rule{16.5cm}{0.4pt}
\parbox{16.5cm}
{\begin{abstract}This paper focuses on the next detectors for gravitational wave astronomy which will be required after the current ground based detectors have completed their initial observations, and probably achieved the first direct detection of gravitational waves. The next detectors will need to have greater sensitivity, while also enabling the world array of detectors to have improved  angular resolution to allow localisation of signal sources. Sect. 1 of this paper begins by reviewing proposals for the next ground based detectors, and presents an analysis of the sensitivity of an 8 km armlength detector, which is proposed as a safe and cost-effective means to attain a 4-fold improvement in sensitivity. The scientific benefits of creating a pair of such detectors in China and Australia is emphasised. Sect. 2 of this paper discusses the high performance suspension systems for test masses that will be an essential component for future detectors, while sect. 3 discusses solutions to the  problem of Newtonian noise which arise from fluctuations in gravity gradient forces acting on test masses. Such gravitational perturbations cannot be shielded, and set limits to low frequency sensitivity unless measured and suppressed. Sects. 4 and 5 address critical operational technologies that will be ongoing issues in future detectors. Sect. 4 addresses the design of thermal compensation systems needed in all high optical power interferometers  operating at room temperature. Parametric instability control is addressed in sect. 5. Only recently proven to occur in Advanced LIGO, parametric instability phenomenon brings both risks and opportunities for future detectors. The path to future enhancements of detectors will come from quantum measurement technologies. Sect. 6 focuses on the use of optomechanical devices for obtaining enhanced sensitivity, while sect. 7 reviews a range of quantum measurement options.
\end{abstract}}
\end{center}\vspace*{-0.6cm}

\begin{center}
\parbox{16.5cm}
{\bf\jiuhao future gravitational wave detectors, opto-mechanics, quantum limit}
\end{center}

\begin{center}
{}
\Cit{Blair D, Ju L, Zhao C N, et al. The next detectors for gravitational wave astronomy. Sci China-Phys Mech Astron,
2015, 58: 120405, doi: 10.1007/s11433-015-5747-7}
\end{center}

\textwidth=178truemm \textheight=236truemm

\wuhao\vspace*{1.5mm}
\tableofcontents
\vspace*{5mm}
\begin{multicols}{2}

\renewcommand{\baselinestretch}{1.08} \baselineskip 12.2pt\parindent=10.8pt

\renewcommand{\thefootnote}


\section{The next  ground based detectors}
\emph{The current generation of gravitational wave (GW) detectors are expected to make initial detections of GW sources. Their signal to noise ratio will be sufficient to catalog the rate of strong sources, and to see them with limited signal to noise ratio. It will be extremely important to develop a next generation of more sensitive detectors that can examine waveforms to enable detailed investigations of black hole and neutron star physics. Here we present a concept for a north-south pair of 8 km arm length detectors.  Such a  system could monitor a volume of the universe 60 times larger than the best current detectors, observing black hole systems in the era of high star formation, as well as observing nearer systems with very high signal to noise ratio.}

\subsection{The need for improved detectors}\label{next detector}

Since Initial LIGO did not observe GW signals during 1 year of observations, it is reasonable to assume that GW signals observed by the current generation of detectors will have limited signal to noise ratio. Signal components such as the ring down of black hole normal modes are likely to be weak, so that detailed testing of general relativistic predictions regarding the physics of black holes is likely to be limited. The next generation of GW detectors is motivated by the need to achieve a significant improvement in sensitivity compared with the existing detectors.   High signal to noise ratio studies of coalescing neutron star binaries should be able to test neutron star physics in an extreme dynamical state. Black hole binaries should be detectable into the medium-high redshift universe, allowing us to probe star formation. The standard siren properties of binary black holes will allow new cosmological studies.

Substantial efforts have been put into designing such detectors. Fundamental to the design of new detectors is the understanding of the noise sources in existing detectors\cite{GWbook}. Thanks to the enormous efforts that have gone into noise diagnosis by the teams developing the first generations of LIGO and Virgo, laser interferometer noise sources are well understood. Designs must take into account seismic noise, thermal noise from various optical coating acoustic loss processes, thermal noise from pendulum suspension losses, quantum noise in the form of radiation pressure noise at low frequency and shot noise at high frequency. Another critical noise source for the future will be gravity gradient noise---short range gravitational force noise due to time varying gravity gradients caused by surface seismic waves and low frequency atmospheric acoustic waves. This noise source is significant for signal frequencies below about 10 Hz.

The Einstein Telescope (ET) design study proposed a triangular arrangement containing up to 6 interferometers of 10 km base line, at least 200 m below ground \cite{4.1ET}. This detector would use one cryogenic test mass system for high frequency detection and a parallel room temperature system with very large test masses for low frequency detection.  The underground location was chosen to reduce gravity gradient  noise.

For several years various options for the next generation detector after Advanced LIGO (aLIGO) have been discussed. One option that has been discussed is a 40 km above ground interferometer. An interferometer of this arm length was first suggested by a Chinese team who even identified a suitable basin for which the major excavation requirements (due to the curvature of the Earth) could be reduced$^{1)}$\footnote{1) CHEUNG Yeuk-Kwan E. Private communication}.

The advantage of a very long arm length is that the strain amplitude of local noise sources associated with the test masses (thermal noise, seismic noise and gravity gradient noise) reduce inversely with the arm length $L$. Thus increased arm length dilutes local noise sources and reduces the technological requirements (such as using cryogenics or reduced loss optical coatings) that are otherwise required for noise suppression.
There are many challenges to this very long arm approach. Firstly,  a large increase in arm length requires the vacuum envelop to be enlarged. Vacuum pipe costs increase as the second or third power of pipe diameter,
because of the strength requirements to prevent collapse. Residual gas dispersion noise increases with arm length so that ultra high vacuum performance is still required.
Mirror figure errors, which already have extremely stringent requirements in Advanced detectors (typically 0.5 nm), must be improved proportional to the arm length because their fractional contribution to radius of curvature error increases proportional to $L$.
Similarly, thermal lensing (which can be eliminated using cryogenics) contributes errors which are proportional to $L$.  Finally there is substantial risk associated with the fact that the arm cavity free spectral range reduces from 37 kHz in aLIGO to 3.7 kHz, which is close to the signal frequency band.
Moreover, the larger diameter test masses, required by diffraction effects, means that the test masses have a very high density of acoustic modes. Under these conditions the problem of parametric instability, due to transverse optical modes scattering resonantly with test mass acoustic modes, is likely to be very severe\cite{zhaoPIPRL}.

During the KITPC Next Detectors for Gravitational Astronomy program there was extensive discussion about a less extreme option: a surface interferometer of 8 km arm length. This design was chosen as a more conservative option that would extend existing technology in step sizes $\sim2$ rather than an order of magnitude. Such an interferometer could be accommodated in beam tubes of 1.2 m diameter (the same as LIGO), and would use test masses of 80 kg, double the mass of LIGO test masses. Suspension thermal noise could be sufficiently reduced by doubling the suspension pendulum lengths, while the thermal noise of the optical coatings could be reduced by averaging the thermal noise over a doubled optical beam size on the test masses. The optical power level needs to increase by only 20\% compared with aLIGO.

To avoid major increase in optical power, it is assumed that 8 dB of frequency dependent optical squeezing can be achieved. This exceeds the best observed noise improvement by 4.3 dB and requires reduced optical losses in the signal recycling cavity. To improve the low frequency sensitivity, a gravity gradient noise suppression system based on both atmospheric pressure monitoring and seismic wave monitoring would be used to suppress gravity gradient fluctuations at low frequency by a factor of 5.

\begin{figure}[H]
\centering
\includegraphics[scale=1]{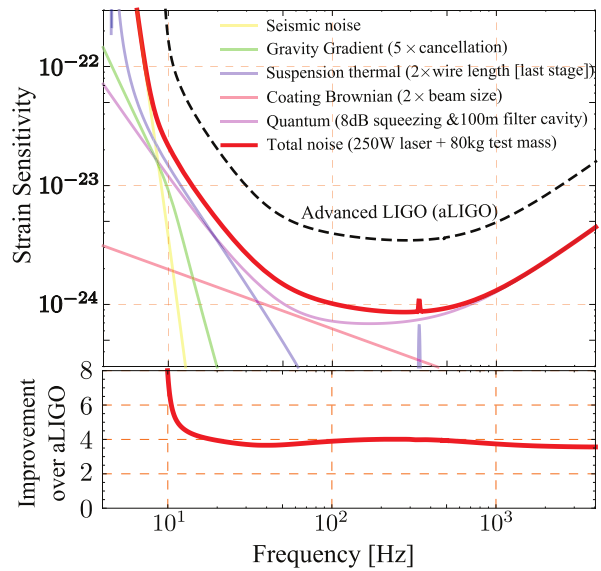}
\vspace*{-2mm}
\caption{(Color online) 8 km Advanced GW detector noise budget. The contributions of various noise sources are indicated in the legend. The improvement factor compared with aLIGO is indicated in the bottom panel. The two-fold scale up of instrument parameters quadruples the instrument sensitivity. } 
\label{fig7_8km}
\end{figure}

As a direct scale-up of aLIGO the 8 km interferometer design would be relatively easily designed and costed. It would be a reliable design with few major unknowns. However, it would still require very careful design to minimise parametric instability, optimise thermal compensation and to optimise the recycling cavities and the mode cleaners. Modelling by Miao shown in Figure \ref{fig7_8km} shows that the 8 km detector could achieve a four fold increase in GW strain sensitivity (equivalent to a 4-fold increase in telescope diameter for an optical telescope). As with optical telescopes, such a step would yield substantially more science, by increasing the horizon range 4-fold and signal event rates 64 times.

Figure~\ref{fig7_8km} shows  the sensitivity of the above 8 km interferometer, compared with aLIGO. The second panel shows the improvement factor as a function of frequency.

For this work we numerically optimized parameters of the interferometer, e.g., the filter cavity detuning and bandwidth, trying to maximize the broadband enhancement over aLIGO.  The same methods were applied in ref.~\cite{haixing}, in which the authors made a systematic comparison of different advanced techniques for enhancing detector sensitivity. The technique of frequency-dependent squeezing, which is included in the above modelling, was shown to be one of the most promising techniques.

In the first paper in this issue it was pointed out that a north-south pair of next generation detectors, one in China and one in Australia,  would improve the array angular resolution from about 14 square degrees to 6 square degrees, thereby creating a global GW telescope with angular resolution reasonably matched to the field of view of electromagnetic telescopes. Because angular resolution improves with signal to noise ratio, the addition of two detectors with the above proposed sensitivity would provide even
 stronger angular resolution benefits.

\subsection{Conclusion}
We have shown that a fourfold  sensitivity improvement can be achieved by doubling several instrument parameters compared with aLIGO, to create an 8 km armlength interferometer. Such a detector design would allow detection of  GW sources into the era of high star formation rates, providing an extremely powerful astrophysical probe of the universe. The next parts of this paper address some of the relevant technologies for this detector.


\section{Ultralow thermal noise suspension systems for ground based gravitational wave detectors}
\emph{The advanced network of GW detectors are on the verge of starting science runs in an era which will see the gravitational window on the universe opened for the first time. The LIGO and VIRGO detectors have undergone significant upgrades which will improve their broadband sensitivity by an order of magnitude, while pushing the lowest operating frequency to 10\;Hz. aLIGO employs a monolithic fused silica final stage suspension in order to lower the thermal noise level to 10$^{\it -19}\; \textrm{m}/\sqrt{\textrm{Hz}}$ at a frequency of 10\;Hz. Fused silica is an ultra-pure high melting point glass which has many desirable properties for a GW suspension. These include exhibiting a mechanical loss approximately 1000 times lower than steel, having the ability to be pulled into long thin suspensions fibres, displaying a breaking stress in excess of 4\;GPa,  and having a Young's modulus which increases with temperature. The latter property allows the suspensions to cancel an otherwise dominant thermoelastic noise component. All the aLIGO monolithic suspensions have now been installed and characterised and are now under vacuum at the Hanford and Livingston sites. There is significant effort focused on warm upgrades to the suspensions. These include changes such as using thinner fibres in order to lower the bounce mode of the suspension and improve the vertical thermal noise. Methods to improve the longitudinal thermal noise performance include pulling fibres from thicker stock and using longer suspensions fibres. Both of these techniques improve the dissipation dilution with a doubling of the suspension improving the dilution by a factor of two. Combined with a heavier test mass (80\;kg) these techniques lead to potential improvements of up to a factor of 3 in suspension thermal noise at 10\;Hz.}

\subsection{Introduction}\label{sec:intro}
There is currently a significant worldwide effort to upgrade the international GW detector network to the 2nd generation status. In Germany the GEO-HF upgrade \cite{GEOHF} to the UK-German GEO600 detector has enabled the instrument to remain the only active instrument during the major upgrades of LIGO and VIRGO. GEO-HF has been operating in Astrowatch mode at over 62 percent duty cycle \cite{dooly}, while demonstrating squeezing on a full scale interferometer to improve the high frequency shot noise level by a factor of approximately 2 \cite{GEO_squeeze}. The LIGO detectors comprise two 4\;km detectors located at Hanford, State of Washington and Livingston, Louisiana, while in Italy the 3\;km VIRGO detector is located near Cascina, Pisa. These detectors are currently being upgraded to aLIGO \cite{aLIGO} and Advanced VIRGO \cite{VIRGO}; involving a major upgrade of the laser system, main interferometer mirrors and the suspension systems.

\begin{figure}[H]
\centering
\includegraphics[scale=0.65]{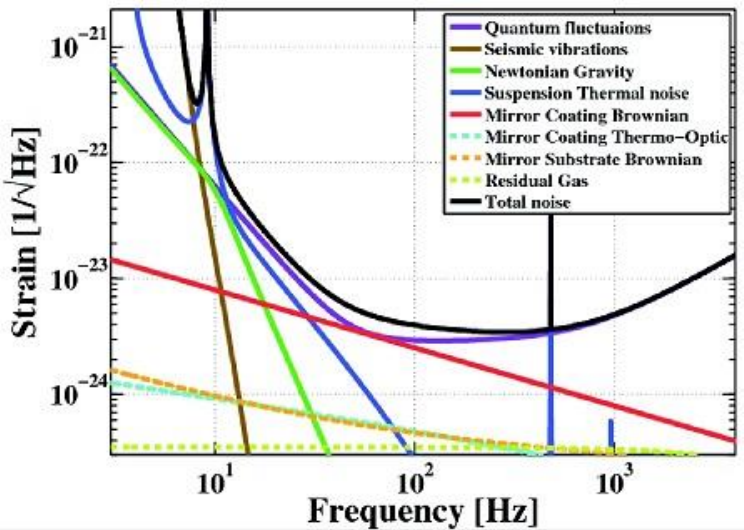}\vspace*{-1mm}
\caption{(Color online) Noise budget for the aLIGO detector.\hspace*{14mm}}
\label{figure1}
\end{figure}

\begin{figure}[H]
\centering
\includegraphics[scale=0.45]{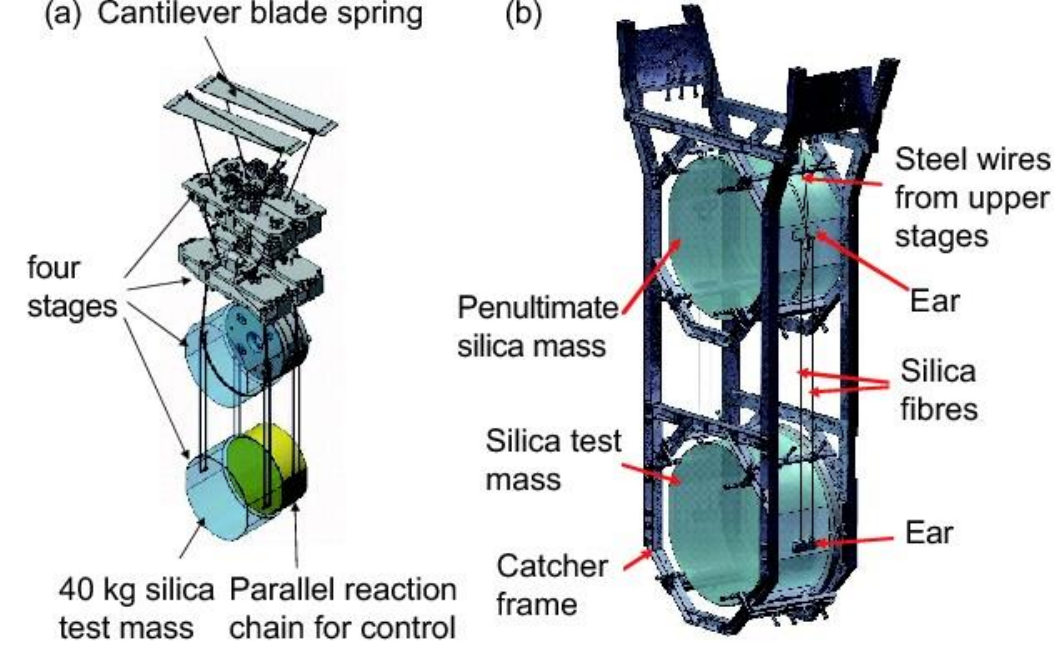}\vspace*{-1mm}
\caption{(Color online) (a) Overview of the aLIGO QUAD pendulum suspension; (b) zoom of the QUAD lower monolithic stage.}
\label{figure2}
\end{figure}

Figure \ref{figure1} shows the typical noise projection for the aLIGO detector. At high frequencies above 100\;Hz shot noise is dominant. This noise results from the statistical fluctuation of power in the interferometer beam and can be reduced by injecting more power into the interferometer arms or applying squeezing at the output port. 
Coating thermal noise is dominant in the mid-frequency region around 30--100\;Hz where the detector has the maximum astrophysical reach. Mechanical loss in the mirror coatings, which are directly sensed by the laser beam, give rise to thermal noise via the Fluctuation-Dissipation theorem \cite{welton}.  Significant research is focused on providing high reflectivity low optical scatter/loss coatings for aLIGO \cite{losscoating,dopeta2O5}. 
At frequencies below 30\;Hz the noise sources are numerous and steep. Improvement within this region requires a coordinated approach whereby sources including seismic noise, suspension thermal noise, gravity gradient noise and radiation pressure noise are all pushed downwards and to the left. Gravity gradient noise arises from the density perturbations induced by seismic waves or acoustic waves \cite{GG}. 
There are currently efforts to characterise the low frequency seismic spectrum at the different detector sites in addition to implementing seismometer arrays to allow subtraction of the effect \cite{driggers}. Radiation pressure noise results from the statistical variation in the arm power and combined with shot-noise it is termed quantum noise \cite{quantumnoise}.
Seismic noise originates from excitation of the main suspension from both environmental and man-made ground vibrations. Environmental origins include the microseismic noise at 160\;mHz, earthquakes and wind driven noise, while man-made sources include vehicles, trains and compressors/motors \cite{seismic}. In aLIGO there are seven isolation stages of the main test mass from the ground. 
An external in-air hydraulic actuator \cite{isi} is used to take out large motions due to daily earth tide variations (up to 0.4\;mm) and microseismic motion. This is followed by an in-vacuum 2-stage active-passive isolation platform providing a factor of 1000 isolation at 10\;Hz \cite{isi}. 
To get to a baseline sensitivity of $10^{-19}\; \textrm{m}/\sqrt{\textrm{Hz}}$ at 10\;Hz requires the 4-stage quadruple pendulum (QUAD) \cite{QUAD}. This pendulum, which has been a major deliverable of the aLIGO UK project team (Universities of Glasgow, Strathclyde, Birmingham and Rutherford Appleton laboratory), provides an isolation of more than 10 million at 10\;Hz. The pendulum, shown in Figure \ref{figure2}, comprises a main
chain of 4 pendulum stages to provide horizontal isolation. Vertical isolation is provided by 3 stages of cantilever blade springs which are built into the upper metal masses. The final 2 stages of the QUAD pendulum are 40\;kg fused silica test \linebreak
\vspace*{-2mm}

\vspace*{-2mm}

\noindent masses of diameter 34\;cm and thickness 20\;cm. To provide a suspension thermal noise performance of $10^{-19}\; \textrm{m}/\sqrt{\textrm{Hz}}$ at 10\;Hz, identical to the seismic noise contribution, requires the use of fused silica fibres which are attached to the side of the test mass to form a quasi-monolithic structure.

As well as providing isolation, the QUAD pendulum also applies local and global control signals. Local control is used for damping the body modes of the individual suspension while global control is used to align the optics to ensure that the correct operating point of the interferometer is maintained. These controls are applied through a quiet reaction chain hanging behind the main optic chain. The control is hierarchical such that larger forces are applied at the upper stages. At the topmost 3 stages a combination of magnet-coil actuators/sensors are utilised, while at the lowest stage an electrostatic drive actuates on the mirror \cite{QUAD}. The design of the QUAD suspension is such that 22 out of the 24 body modes can be sensed and actuated at the top stage. The other 2 modes, bounce and roll of the suspension, are not expected to be significantly rung up during normal science mode operation.

\subsection{Monolithic fused silica suspensions}
Thermal noise is described by the fluctuation-dissipation theorem \cite{welton}. The theorem states that a dissipative system in thermal equilibrium will have a corresponding fluctuation in its measurable parameters. A classic example is Johnson noise in a resistor whereby the dissipative resistance gives rise to a fluctuating voltage. In a mechanical pendulum the dissipation gives rise to thermal displacement noise via
\begin{equation}\label{eq1}
S_{x}\left(\omega \right)=\frac{4 k_{\rm B}T}{\omega^{2}} \left(\Re{\left[Y\left(\omega \right)\right]} \right),
\end{equation}
where $S_{x}$ is the displacement power spectral density, $\omega$ is the angular frequency,  $k_{\rm B}$ is the Boltzmann constant, $T$ is the temperature and $Y\left(\omega \right)$ is the mechanical admittance. Applying this to a mechanical pendulum systems gives rise to a set of sharp resonances occurring at the body modes of the suspension. Examples include the longitudinal and bounce modes which occur at 0.6\;Hz and 9.4\;Hz respectively for aLIGO. Above the final body mode of the suspension the displacement noise can be shown to be \cite{gonzalez}
\begin{equation}
S_{x}\left(\omega \right)=\frac{4 k_{\rm B}T \omega_{0}^{2} \phi \left(\omega \right)}{M \omega^{5}},
\end{equation}
where $M$ is the mass of the pendulum and $\phi \left(\omega \right)$ is the mechanical loss. The equipartition theory states that the integral of the total energy stored in the potential or kinetic terms should be
$\frac{1}{2}k_{\rm B}T$. Ultra-low loss materials store the majority of the thermal energy at the body mode resonances, thus lowering the off-resonance thermal noise. 
The operating bandwidth of the detector is thus in between the suspension resonances, at frequencies $<$10\;Hz, and the internal modes of the mirror at frequencies $>$few kHz.

\subsubsection{Mechanical loss terms}
Fused silica is high melting point glass which displays a mechanical loss which is a factor of 1000 lower than steel. The Suprasil family of glasses are used throughout the aLIGO detector with 3001 for the input test masses and 311/312 for the end test masses, and Suprasil 2 for the fibres. There are several sources of mechanical dissipation in a loaded fused silica suspension fibre \cite{penn}. These include surface loss which is dependent on the ratio of the surface area to volume of the fibre. Empirical measurements show that fibres exhibit a lossy surface, possibly due to surface damage or micro-cracks, following $\phi_{\rm surface}=8 h\phi_{s}/d$, where $h\phi_{s}\simeq 6\times 10^{-12}$ for fused silica and $d$ is the fibre diameter. The attachment of the fibres is done via laser welding with a carbon dioxide (CO$_{2}$) laser. This results in a small region with a higher than nominal thermal stress which results in  a weld loss term in the mechanical loss model. Measurements on fibres have shown that this term can vary on the geometry and quality of the weld but is typically at the level $\phi_{\rm weld}=5.8\times 10^{-7}$ \cite{heptonstall}. The final important mechanical loss term is thermoelastic loss. This arises from the coupling of statistical temperature fluctuations via the thermo-mechanical properties of the suspension fibres. Consider for example a fibre with a thermal expansion coefficient $\alpha$. When this fibre undergoes bending one side of the fibre contracts while one side expands. This sets up a heat flow across the fibre resulting in dissipation. The theory describing thermoelastic loss was first developed by Zener \cite{zener}. More recently it was realised that the temperature variation of the Young's modulus, or $\beta =1/Y \left[\d Y/\d T\right]$, also plays a key role in defining thermoelastic loss in loaded suspension fibres \cite{cagnoli}. The full expression for thermoelastic loss is thus given by
\begin{equation}
\phi_{\rm thermoelastic}=\frac{Y T}{\rho C}\left(\alpha-\sigma_{o}\frac{\beta}{Y}\right)^2\left(\frac{\omega\tau}{1+(\omega\tau)^2}\right),
\end{equation}
where $C$ is the specific heat capacity, $\sigma$ is the stress in the fibre and $\tau$ is the characteristic time for heat to flow across the fibre. It is convenient to consider $\left(\alpha-\sigma_{o}\frac{\beta}{Y}\right)$ as the effective thermal expansion coefficient. All metals display a Young's modulus that reduces as the temperature increases, however, fused silica has a Young's modulus which increases as the temperature increases. As a result $\beta$ has a positive value for fused silica and allows for the possibility of canceling the effective thermal expansion coefficient. This is a remarkable result given that all that needs to be done is to change the stress in the fibre by choosing an appropriate geometry. This has recently been demonstrated in an experiment to measure the thermal expansion coefficient in a loaded silica fibre where it was shown that at a stress of $\simeq 175$\;MPa \cite{bellDC} the silica fibre showed no expansion/contraction under a variable heating.

The most robust thermal noise models of aLIGO suspensions have been developed by profiling the real fibres used in the suspensions and generating a Finite Element model in ANSYS \cite{cumming2}. At each point along the fibre the geometry can be utilised to predict the surface, weld and thermoelastic loss terms; $\phi_{\rm fibre}=\phi_{\rm weld}+\phi_{\rm surface}+\phi_{\rm thermoelastic}$. This loss is then scaled by the energy distribution in the fibre \cite{cumming2} as regions which store no strain energy do not contribute to the total mechanical loss. For the longitudinal mode it is the ends of the fibres which are critical for the low thermal noise operation of the suspension. It is further important to note that loaded suspension fibres have an additional term which dilutes, or reduces, the mechanical loss. A thin suspension fibre stores energy in both elastic bending and gravity. Elastic bending results in a loss, $\phi_{\rm fibre}$, while gravity is a conservative force and does not exhibit any loss. Thus as the majority of the energy is stored in gravity the total fibre mechanical loss, which appears in the expression for the fluctuation-dissipation theorem, is diluted by the ration of $D=E_{\rm gravity}/E_{\rm elastic}\simeq E_{\rm gravity}/E_{\rm elastic}$. Again ANSYS finite element modeling can be used to generate the most robust dilution estimates which account for the real fibre geometry and the transition from fibre to a thicker, but not infinitely rigid, attachment point. For aLIGO the dilution is approximately 90 \cite{cumming1}.
\subsubsection{Suspension fabrication}
There has been significant development of the techniques necessary to provide a robust suspension for aLIGO. The GEO600 detector pioneered the use of fused silica suspension technology which was then transferred to aLIGO via expertise at University of Glasgow. The fibres used in aLIGO are 60\;cm long and 400\;$\upmu\textrm{m}$ at their thinnest section. The thin section is to ensure that the bounce mode of the suspension is $<$10\;Hz while the lowest violin mode is $>$450\;Hz \cite{cumming1}. The fibres need to thicken to 800\;$\upmu\textrm{m}$ at either end in order to set the stress to the thermoelastic cancellation value as previously described. The weld points are provided by fused silica ears which are hydroxide catalysis bonded onto the side of the test mass \cite{marielle}. The bonds ensure a strong connection with low thermal noise performance and are essential to build a quasi-monolithic suspension.

The fibres are produced from Suprasil 2 stock which has an initial diameter of 3\;mm. The fibre is drawn down using a feed-pull method with heating from a CO$_{2}$ laser operating at 10.6\;$\upmu\textrm{m}$ \cite{CO2}. The laser pulling method ensures a well defined fibre geometry which is free of any contaminants which may be induced by a flame pulling process. The fibres which are produced are pristine and exhibit tensile strengths in excess of 4\;GPa \cite{kirill}. This means that an aLIGO fibre can suspend a load of approximately 65\;kg before breaking. As the four suspension fibres each carry a load of 10\;kg for the 40\;kg optic there is a comfortable safety margin. There are three laser pulling machines that have been built by Glasgow. One is at Glasgow for ongoing machine development while the other two systems are located in Cascina and Hanford for Advanced VIRGO and aLIGO suspension construction.

A portable CO$_{2}$ laser system has been built at both the Hanford and Livingston sites for the purpose of suspension welding, while the single pulling machine in Hanford produces all fibres which are then shipped to Livingston in dry nitrogen storage containers. Prior to installation all fibres are profiled with a non-contact optical profiler to measure their geometry for thermal noise modelling \cite{cumming3}. The fibres are also proof tested at a load of 15\;kg for 5 min in order to ensure that no damage has occurred to the fibres during transport. The fibres are welded into the suspension structure using a custom set of tooling and procedures \cite{cumming1}. The welding process has been transferred to members of the aLIGO suspension team via dedicated training sessions both in the UK and US. Over 20 individual test suspensions and final article suspensions have now been performed for aLIGO. During the final article suspension construction a 100 percent success rate was observed during the fibre welding and hanging procedure. 

An important part of the aLIGO installation and commissioning scheme is phased testing. This starts at the suspension hanging stage where all six body modes, the pitch of the optic and the first violin modes of each of the four loaded suspension fibres are measured in air. These preliminary tests provide an initial confirmation that the suspension is within specifications. These include a pitch difference between the penultimate mass and test mass 
of no more than 2\;mrad, a bounce mode $<$10\;Hz and violin modes $>$\;450 Hz. There is a procedure to adjust the pitch of the suspension \cite{cumming1} if required although this has only been utilised once during the entire installation procedure. Furthermore we find that the violin modes are clustered with $\pm1$ percent which is further confirmation that the load is equally distributed and the fibres are of similar geometry. The suspension/isolation system characterisation procedure then proceeds to in-chamber testing both in air and finally in-vacuum. Such a procedure has enabled an accurate model of each individual suspension to be developed in addition to significantly speeding up commissioning activities.
\subsubsection{Suspension thermal noise}
Refs. \cite{cumming2, cumming1} provide a detailed analysis of the thermal noise performance of the suspension. The main contributions are from the longitudinal and bounce modes of the suspension 
At 10\;Hz the noise is a factor of two lower than the design sensitivity of $10^{-19}\; \textrm{m}/\sqrt{\textrm{Hz}}$.

Work is currently underway to measure the quality factor ($Q$) of the violin mode ringdowns as these given an indication of the loss mechanisms present in the fibres. Although measured at $\simeq500$\;Hz they are a useful method to identify any excess loss terms. The challenge is that the $Q$ of these modes is approaching 1 billion ($\phi\simeq 1\times10^{-9}$) resulting in a ringdown of several days. With the current commissioning schedule, leading up to the first science run in late 2015, it is challenging to get long periods of time to perform these measurements. However, $Q$ factors of up to 700 million are predicted by the finite element model assuming the nominal values for weld loss and surface loss. This is a very positive result and seems consistent with measurements, and it is possible that the weld loss term, which shows most variability, could be lower due to improved welding techniques at the sites. This can be verified by monitoring several violin mode losses and fitting to individual loss terms at a later stage.

\subsection{Suspension upgrades}
There is still a significant effort focused on future warm upgrades to the suspensions. These can roughly be split into those requiring minor suspension modification and those requiring more substantial changes. Changes of the former  include using thicker silica stock to pull the fibres (5 mm rather than 3 mm) and using thinner suspension fibres \cite{hammond, heptonstall2}. A thicker end stock improves the dilution by realising an attachment which is more rigid, thus reducing the effect of weld loss at the fibre ends. Operating at a higher stress such as 1.5\;GPa in the thin section (comapred to the current 800\;MPa) would lower the bounce mode of the suspension from $\simeq 9.4$\;Hz to $\simeq 6$\;Hz. These changes can improve thermal noise performance at 10\;Hz by a factor of 1.5.

More substantial changes, requiring redesign of suspension components, include utilising longer suspension fibres (1.2\;m rather than 60\;cm) and heavier test masses (80\;kg rather than 40\;kg) in addition to the option of thicker stock and thinner fibres noted above. Such a suspension would have a further reduction in the bounce mode and also a $\times 2$ improvement in the dilution, as this term is inversely proportional to the fibre length \cite{cumming1}. Thus we could expect a suspension thermal noise which is more than three times better than current aLIGO at 10\;Hz. Work is currently underway to assess the feasibility of such a design via full scale testing in Glasgow, in addition to working with aLIGO colleagues to assess the impact of such changes to the suspension and seismic performance.
\subsection{Summary and conclusion}
All of the ultra-low noise monolithic suspensions have now been installed at the aLIGO detector sites of Hanford and Livingston. Fused silica is the material of choice for these suspensions as it has a mechanical loss approximately 1000 times lower than steel, can be pulled into long thin fibres and welded to fused silica attachment points, and displays a breaking stress in excess of 4\;GPa. Fused silica further has the unique property that its Young's modulus increases with increasing temperature, allowing fibres to be constructed with zero effective thermal expansion coefficient, thus minimising the themoelastic noise contribution.

There is robust set of techniques to fabricate, install and characterise these suspensions which are essential to meet the thermal noise target of $10^{-19}\; \textrm{m}/\sqrt{\textrm{Hz}}$ at 10\;Hz. Work is currently underway to characterise the mechanical loss of the violin modes of the suspension in vacuum, while a set of warm upgrade options could allow for a further 1.5--3 improvement in the thermal noise performance at 10\;Hz.


\section{Newtonian noise reduction: Low frequency mechanical detectors and mitigation of Newtonian noise}
\subsection{Introduction}\label{sec:intro}

The Newtonian gravity noise (NN) generated by moving local masses poses a formidable challenge to detection of GWs below 1 Hz. No GW detector can be shielded from the NN.

According to general relativity, a GW is a {\it transverse} wave, whereas a near-field gravity gradient generally has longitudinal components. Therefore, a {\it tensor} detector, which measures all the components of the curvature tensor, can in principle discriminate and reject the NN. In reality, superposition of many seismic and atmospheric waves complicates the rejection procedure. We discuss procedures of subtracting the NN from the GW channels of a tensor detector and discuss their limits for Rayleigh waves and infrasound waves. We also discuss the possibility of mitigating the NN from advanced laser interferometers by directly detecting and removing the NN with tensor detectors co-located with the interferometer test masses.

\subsection{Full-tensor gravitational wave detectors}

According to general relativity, a gravitational field is characterized by a curvature tensor. Terrestrial laser-interferometer GW detectors measure only one off-diagonal component by combining two orthogonal light cavities. A full-tensor detector could be constructed by measuring five degenerate quad-rupole modes of a solid sphere \cite{1Wagoner,2Johnson}. A tensor detector is sensitive to GWs coming from
\emph{any direction with any polarization} and is thus capable of resolving the source direction and polarization.

One could construct a low-frequency (0.01--10 Hz) tensor GW detector by using six almost free test masses \cite{3Paik}. Figure~\ref{f1-model} shows the test mass configuration of such a detector, named SOGRO (Superconducting Omni-directional Gravitational Radiation Observatory). The entire detector is
cooled to 1.5 K by pumping on liquid helium or by using cryocoolers. Six superconducting test masses,
each with three linear degrees of freedom, are levitated over three orthogonal mounting tubes.
The test masses are made of niobium (Nb) in the shape of a rectangular shell. Superconducting
levitation/alignment coils and sensing capacitors (not shown) are
 located in the gap between the
test masses and the mounting tubes, as well as on the outer surfaces of the test masses.

The along-axis motions of the two test masses on each coordinate axis are differenced to
measure a diagonal component of the wave:

\begin{figure}[H]
\centering
\includegraphics[scale=0.8]{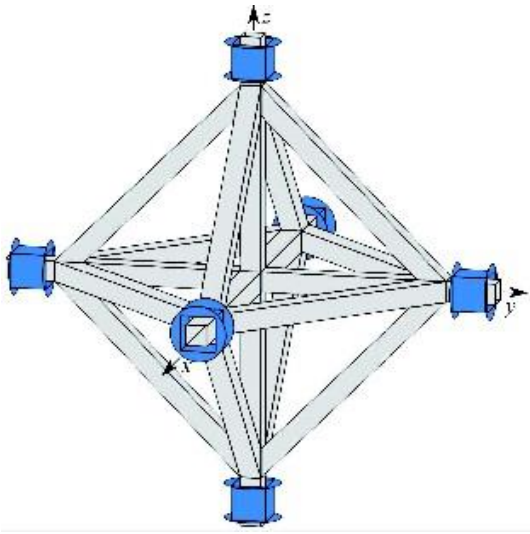}\vspace{-3mm}
\caption{(Color online) Test mass configuration for SOGRO. Motions of six magnetically levitated test masses are combined to measure all six components of the curvature tensor.}
\label{f1-model}
\end{figure}

\begin{equation}\label{eq1}
h_{ii}(t)=\frac{2}{L}[x_{+ii}(t)-x_{-ii}(t)],\vspace*{-1mm}
\end{equation}
where $x_{\pm ij}(t)$ is displacement of the test mass on the $\pm i$ axis along the $j$-th axis and $L$ is the separation
between the test masses on each axis. The cross-axis (rotational) motions of the four test
masses on each coordinate plane are differenced to measure an off-diagonal component of the
wave:
\begin{equation}\label{eq2}
h_{ij}(t)=\frac{1}{L}\{[x_{+ij}(t)-x_{-ij}(t)]-[x_{-ji}(t)-x_{+ji}(t)]\}, i\neq j.
\end{equation}

In addition to measuring the six strain signals, the detector will measure the three linear and three
angular platform acceleration signals by \emph{summing} the along-axis and cross-axis test mass motions:
\begin{equation}\label{eq3}
a_{ii}(t)=-\frac{1}{2}\omega^{2}[x_{+ii}(t)+x_{-ii}(t)],
\end{equation}
\begin{equation}\label{eq4}
\alpha_{ij}(t)=-\frac{1}{2L}\omega^{2}\{[x_{+ij}(t)-x_{-ij}(t)]+[x_{-ji}(t)-x_{+ji}(t)]\}, i\neq j.
\end{equation}
These common-mode (CM) acceleration signals are used to remove the residual sensitivity of the
differential-mode (DM) channels of the detector to the platform accelerations \cite{4Moody}.
Since test mass motion is measured with respect to the sensing circuit elements mounted on
the platform, this detector requires a rigid platform with mode frequencies above the signal
bandwidth, $> 10$ Hz. To reduce its thermal noise, the platform itself needs to be cooled to 1.5 K.
The design detail of SOGRO, including a description of the readout transducer and an analysis of
the detector noise, will be published elsewhere \cite{3Paik}.

\subsection{Mitigation of Newtonian noise on SOGRO}

At low frequencies,
the NN is dominated by Rayleigh waves and infrasound waves. For a laser interferometer
to overcome this noise below 1 Hz, the ground and air motion within tens of kilometers from the
detector must be measured with a large number of seismometers and microphones with sufficient
accuracy, and then the induced NN computed and subtracted from the detector output. The NN
from Rayleigh waves could be canceled up to one part in $10^3$ by using this method \cite{5Harms}. But for infrasound waves, cancellation works only for waves coming \textit{in certain favorable directions}, as
we will see.

In contrast, by using its tensor nature, SOGRO can mitigate the NN from both Rayleigh
waves and infrasound waves to one part in $10^3$ for \textit{all incident angles}. A detailed analysis of NN
mitigation for SOGRO has been published elsewhere \cite{6Harms2}. Here we summarize the result.

Assuming that the interferometer is underground at depth $z < 0$, the gravitational perturbation
of a single test mass due to a Rayleigh wave incident at an angle $\psi$ with respect to the sensitive
axis $x$ of the test mass and an infrasound wave incident in direction ($\psi , \vartheta$) is given \cite{5Harms} by
\begin{align}\label{eq5}
X(\omega)=&-2\pi {i}\cos{\psi} G\rho_0\gamma_{\rm R}\frac{\xi(\omega)}{\omega^2}{\rm exp}\left(\frac{\omega}{c_{\rm R}}z\right)\nonumber\\
&-4\pi G\sin\vartheta\cos\psi\frac{\delta\rho(\omega)c_{\rm IS}}{\omega^3}{\rm exp}\left(\frac{\omega}{c_{\rm IS}}z\sin\vartheta\right),
\end{align}
where $\xi(\omega)$ and $\delta\rho(\omega)$ are the vertical ground displacement and atmospheric density fluctuation
directly above the test mass, $\gamma_{\rm R} \approx 0.83$ is a factor that accounts for the partial cancellation for the
Rayleigh NN from surface displacement by the sub-surface compressional wave content of the
wave field, $c_{\rm R} \approx 3.5$ km/s and $c_{\rm IS} \approx 330$ m/s are the speed of the Rayleigh waves underground and
the infrasound waves, respectively, and $\rho_0$ is the mean mass density of the ground. The metric
perturbation tensor \textit{in the detector coordinates} can be shown to be
\begin{align}\label{eq6}
&h_{NG}(\omega)=\nonumber\\
&\left[2\pi G\rho_0\frac{\gamma_{\rm R}}{c_{\rm R}}\frac{\xi(\omega)}{\omega}{\rm exp}\left(\!
\frac{\omega}{c_{\rm R}}z\!\right)\!+\!4\pi G \frac{\delta \rho(\omega)}{\omega^2}
\sin^2\vartheta {\rm exp}\left(\!
\frac{\omega}{c_{\rm IS}}z\sin\theta\right)\!\right]\nonumber\\
&\times\left(
\begin{array}{ccc}
\cos^2\psi&\cos\psi\sin\psi&-{\rm i}\cos\psi\\
\cos\psi\sin\psi&\sin^2\psi&-{\rm i}\sin\psi\\
-{\rm i}\cos\psi&-{\rm i}\sin\psi&-1\\
\end{array}\right).
\end{align}

Consider a GW coming from ($\theta, \phi$) direction in the presence of multiple Rayleigh and infrasound
waves. The full strain tensor \textit{in the GW coordinates} has the form:
\begin{align}\label{eq7}
&h'(\omega)=\nonumber\\
&\left(
\begin{array}{ccc}
h_+(\omega)+h'_{NG,11}(\omega)&h_\times(\omega)+h'_{NG,12}(\omega)&h'_{NG,13}(\omega)\\
h_\times(\omega)+h'_{NG,12}(\omega)&h_+(\omega)+h'_{NG,22}(\omega)&h'_{NG,23}(\omega)\\
h'_{NG,13}(\omega)&h'_{NG,23}(\omega)&h'_{NG,33}(\omega)
\end{array}
\right).
\end{align}

Due to the \textit{transverse nature} of the GW, $h'_{13}, h'_{23}$ and $h'_{33}$ contain only the NN components.
Therefore, to recover $h_{+}(\omega)$ and $h_\times(\omega)$, the NN could be removed from $h'_{11}$ and $h'_{12}$ by correlating
them with $h'_{13}, h'_{23}$ and $h'_{33}$, and possibly also with some CM channels and subtracting the
correlated parts.

By combining the tensor components, we find
\begin{subequations}
\begin{align}
\begin{split}
&h_{+}(\omega)=h'_{11}(\omega)-2\cot\theta h'_{13}(\omega)+\cot^2\theta h'_{33}(\omega)\\
&\quad\qquad+\csc^2\theta\frac{2\pi G\rho_0}{\omega}\frac{\gamma_{\rm R}}{c_{\rm R}}{\rm exp}\left(\frac{\omega}{c_{\rm R}}z\right)\sum_i\xi_i(\omega)\\
&\quad\qquad+\csc^2\theta\frac{4\pi G}{\omega^2}\sum_i\delta\rho_i(\omega)\sin^2\vartheta_i{\rm exp}\left(\frac{\omega}{c_{\rm IS}}z\sin\vartheta_i\right),
\label{eq8a}
\end{split}\\
\begin{split}
&h_\times(\omega)=h_{12}'(\omega)-\cot\theta h_{23}'(\omega)\\
&~-{i}\csc\theta\frac{2\pi G\rho_0}{\omega}
\frac{\gamma_{\rm R}}{c_{\rm R}}{\rm exp}\left(\frac{\omega}{c_{\rm R}}z\right)\sum_i\xi_i(\omega)\sin(\psi_i-\phi)\\
&~-{i}\csc\theta\frac{4\pi G}{\omega^2}\sum_i\delta\rho_i(\omega)\sin^2\vartheta_i{\rm exp}\left(\frac{\omega}{c_{\rm IS}}z\sin\vartheta_i\right)\sin(\psi_i-\phi).
\label{eq8b}
\end{split}
\end{align}
\end{subequations}

\begin{figure}[H]
\centering
\includegraphics[scale=0.45]{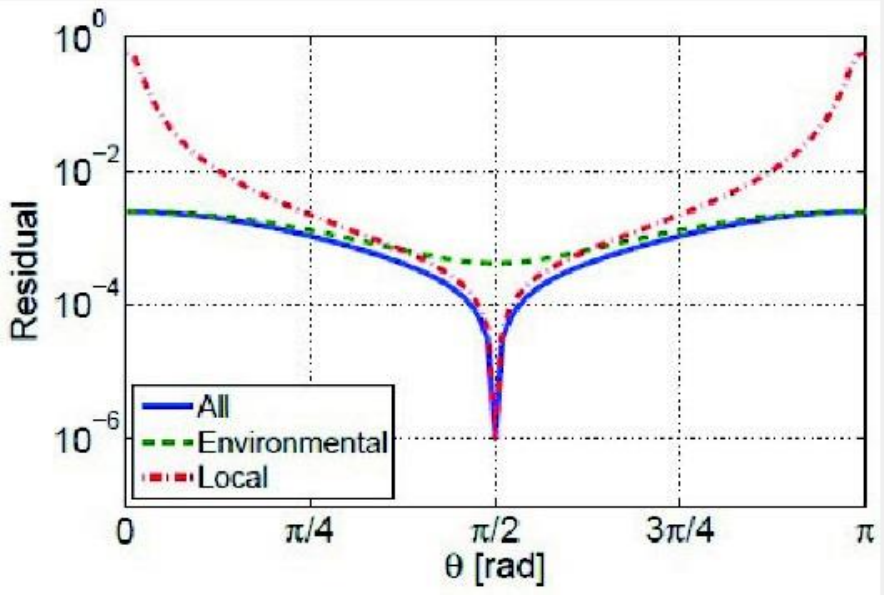}\vspace{-2mm}
\caption{(Color online) NN due to Rayleigh waves removed to $\sim 10^{-3}$ by
using $h'_{13}, h'_{23}, h'_{33}$ and $a_z$ (vertical CM), plus seven
seismometers with SNR = $10^3$ at the radius of 5 km.}
\label{f2-NN}
\end{figure}

Figure~\ref{f2-NN} shows the residual NN achieved for Rayleigh waves \textit{in the absence} of infrasound
waves by using $h'_{13}, h'_{23}, h'_{33}$ and $a_z$, plus seven seismometers with signal to noise ratio
of $10^3$ at the radius of 5 km as the input of the Wiener filter. The NN has been removed to about
$10^{-3}$ with environmental sensors (seismometers) alone. The local channels of SOGRO improve
the noise significantly only near $\theta = \pi/2$, where the noise of the DM and CM channels drop out
according to eq.~(\ref{eq8a}). A tensor detector does not have much advantage over a single-component
detector like the laser interferometer for removing the NN due to the Rayleigh waves.

\begin{figure}[H]
\centering
\includegraphics[scale=0.97]{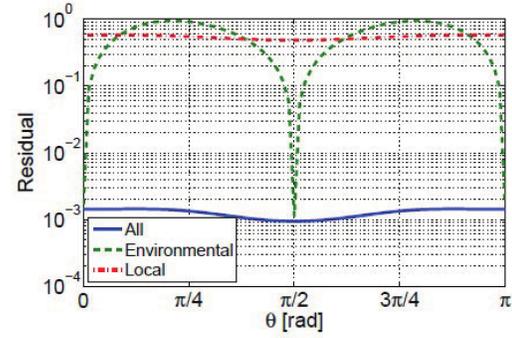}\vspace{-2mm}
\caption{(Color online) NN due to infrasound removed to $\sim 10^{-3}$ by
using $h'_{13}, h'_{23}, h'_{33}$ and 15 microphones of SNR = $10^4$, one at
the detector, seven each at the radius of 600 m and 1 km.}
\label{f3-NN}
\end{figure}

Figure~\ref{f3-NN} is the residual NN achieved for infrasound waves \textit{in the absence} of Rayleigh waves
by using $h'_{13}, h'_{23}, h'_{33}$ and 15 microphones of signal to noise ratio of $10^4$, one at the detector, seven each at the
radius of 600 m and 1 km around the detector. With the environmental sensors (microphones)
alone, the NN \textit{cannot} be mitigated except at $\theta = 0, \pi/2$ and $\pi$. This is because the infrasound
waves come from a half space and the microphones deployed over a surface is insufficient to
measure the effect of 3D density variations of the atmosphere. Thus mitigation of infrasound
NN constitutes a formidable challenge for laser interferometers. In SOGRO, the vertical strain
component $h'_{33}$ largely makes up for this deficiency. With the aid of the local strain channels,
the NN has been rejected to $10^{-3}$ \emph{for all} $\theta$.

\textit{With both waves present}, the resulting \textit{mixed} NN cannot be removed by using the above
methods since the detector channels cannot distinguish NN from the Rayleigh and infrasound
waves. Instead, we can use an array of external seismometers to first remove the Rayleigh waves
\textit{independently of}  the infrasound to one part in $10^3$, and then combine \textit{cleaned-up} $h'_{13}, h'_{23}, h'_{33}$
and microphone signals to remove the remaining
NN due to infrasound to one part in $10^3$. This
will satisfy the rejection requirement for both types of NN for SOGRO\linebreak \cite{3Paik}.

\subsection{Mitigation of Newtonian noise on interferometers with the aid of SOGROs}

Since SOGRO is a very sensitive gravity strain gauge, one may be able to employ scaled-down
SOGROs, in place of a large array of seismometers, to directly measure and remove the
NN affecting the interferometer test masses. Here we describe a procedure of mitigating the NN
in interferometers by using mini-SOGROs with arm-length $l\ll L$ and investigate its limit. We
restrict our discussion to underground detectors like KAGRA \cite{7Somiya} or ET \cite{8ET}.

\begin{figure}[H]
\centering
\includegraphics[scale=0.4]{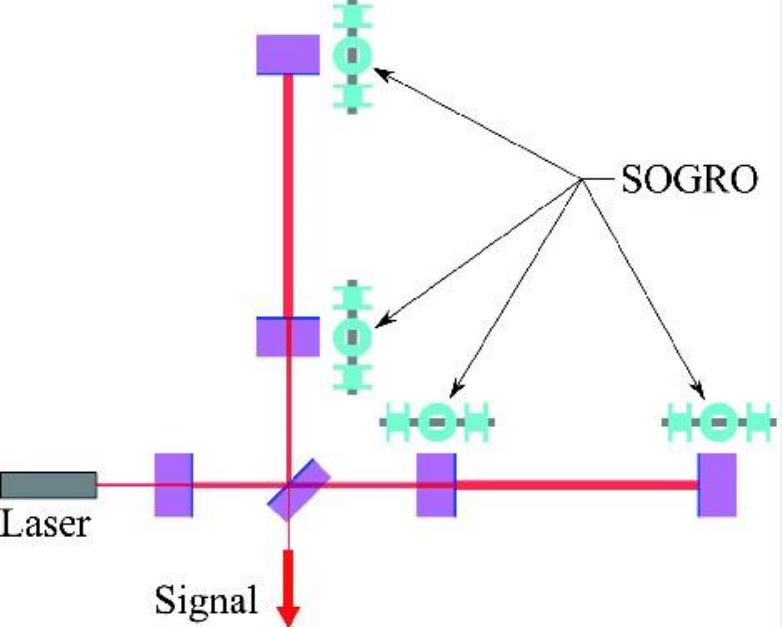}\vspace{-2mm}
\caption{(Color online) Four mini-SOGROs co-located with four test masses
of a laser interferometer GW detector.}
\label{f4-interfer}
\end{figure}

The Rayleigh waves are expected to dominate the NN for an underground detector \cite{9Beker}. In
the presence of a GW and Rayleigh waves, the arm-length along the $x$ axis is modulated by
\begin{equation}\label{eq9}
\Delta L=\frac{hL}{2}+X_{\rm R}(x_{2})-X_{\rm R}(x_{1}),
\end{equation}
where $X_{\rm R}(x_{i})$ is the first term of eq.~(\ref{eq5})
summed over multiple waves for the $i$-th
test mass on the $x$ axis. At 10 Hz, the Rayleigh
wavelength becomes $\lambda_{\rm R} \sim 350~ {\rm m} \ll
L$, causing $X_{\rm R}(x_{1})$ and $X_{\rm R}(x_{2})$ to be uncorrelated.
Hence we measure $X_{\rm R}(x_{i})$ for
each test mass by using a separate SOGRO
co-located with it, as shown in Figure \ref{f4-interfer}.

From eqs.~(\ref{eq5}) and ~(\ref{eq6}), we find that the
13-component of the SOGRO output is
related to $X_{\rm R}(x_{i})$ by
\begin{equation}\label{eq10}
\eta_{13}(x_{i})=h_{13}+2X_{\rm R}(x_{i})\frac{{i}\omega}{c_{\rm R}}.
\end{equation}
We solve eq.~(\ref{eq10}) for $X_{\rm R}(x_{i})$ and substitute it into eq.~(\ref{eq9}) to obtain
\begin{equation}\label{eq11}
h=\frac{2\Delta L}{L}-{i}\frac{c_{\rm R}}{\omega L}[\eta_{13}(x_{2})-\eta_{13}(x_{1})].
\end{equation}
The sensitivity required for mini-SOGRO to recover $h$ is then given by
\begin{equation}\label{eq12}
\eta=\frac{2\Delta l}{l}=\frac{1}{\sqrt{2}}\frac{\omega L}{c_{\rm R}}h.
\end{equation}

\begin{figure}[H]
\centering
\includegraphics[scale=0.5]{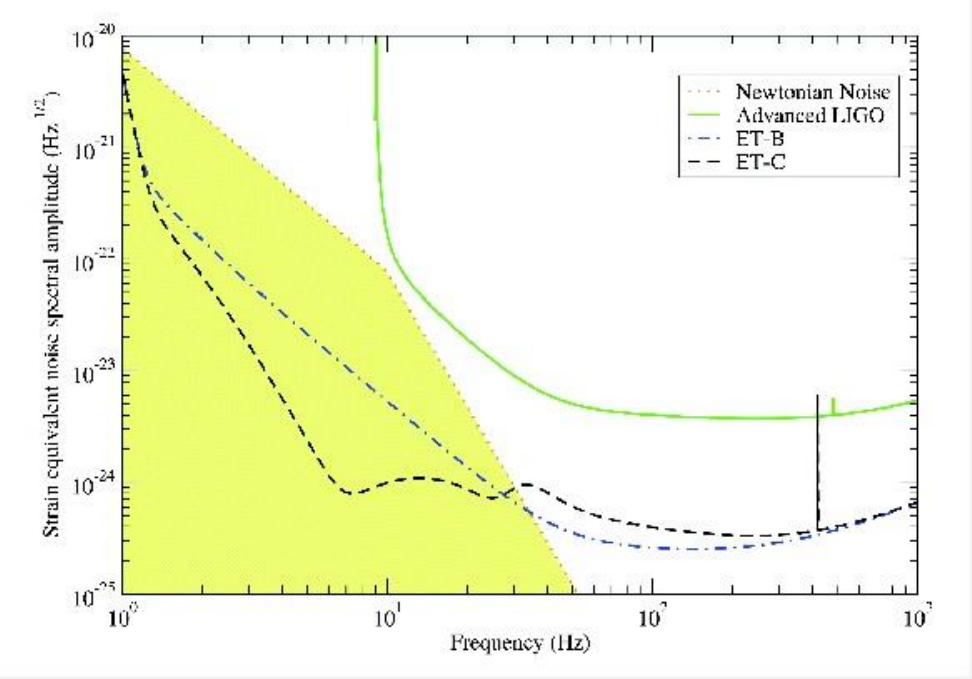}\vspace{-3mm}
\caption{(Color online) Sensitivity goals of aLIGO and ET. The shaded
region represents the parameter space dominated by the NN. Figure 8 is reproduced from Figure 1 in ref. [41].}
\label{f5-sens}
\end{figure}

Figure~\ref{f5-sens} shows the sensitivity goals
of aLIGO and ET \cite{9Beker}. 
The shaded
region represents the parameter space
dominated by the NN. A worthy mitigation
goal would be rejecting the NN
by an order of magnitude to $h \approx 10^{-22}
~{\rm Hz}^{-1/2}$ at 3 Hz and to $10^{-23}$ Hz$^{-1/2}$ at 10 Hz. For ET with $L = 10$ km, eq. ~(\ref{eq12})
yields $\eta\approx4\times 10^{-21}$ Hz$^{-1/2}$ at 3 Hz and
$1.3 \times 10^{-21}$ Hz$^{-1/2}$ at 10 Hz. The NN
between SOGRO test masses must be
highly correlated. According to Beker et
al. \cite{9Beker}, the mitigation factor $S$ is given by
\begin{equation}\label{eq13}
S=\frac{1}{\sqrt{1-C^{2}_{SN}}}\leq\frac{c_{\rm R}}{\omega l},
\end{equation}
where $C_{SN}$ is the correlation between the
test masses. To obtain $S = 10$ at 10 Hz,
we need $C_{SN} = 0.995$ and $l\leq c_{\rm R}/\omega S = 5.6$
m. Mitigating the NN for ground detectors
is more challenging since the
low speed of the Rayleigh waves on the surface, $c_{\rm R} \approx 250$ m/s reduces $l$ to $\leq 0.4$ m. Such a small
SOGRO would hardly have enough sensitivity.

\begin{figure}[H]
\centering
\includegraphics[scale=0.4]{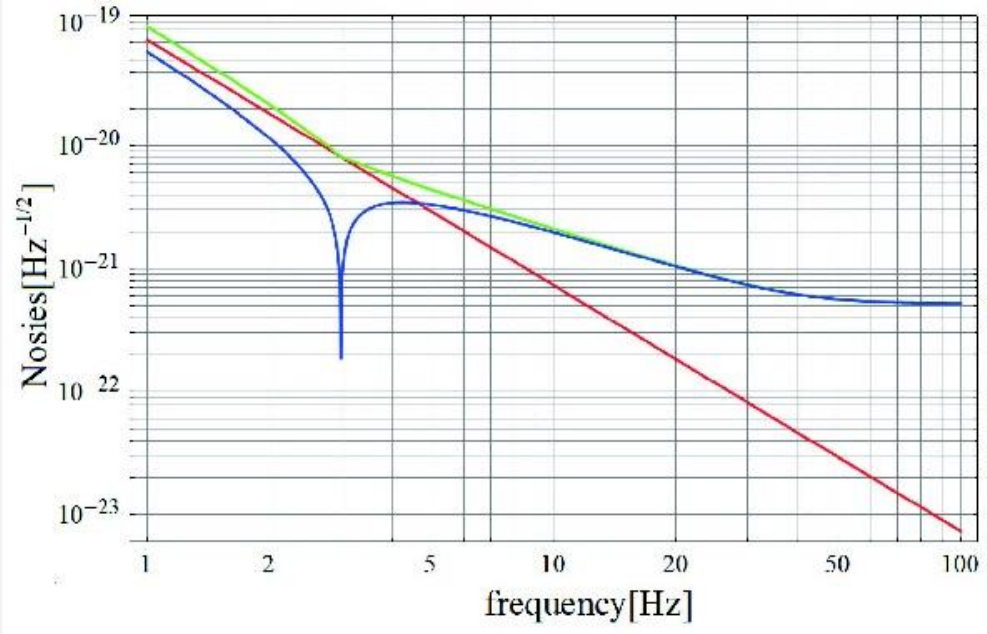}\vspace{-3mm}
\caption{(Color online) Instrument noise spectral density of mini-SOGRO of 5 m arm-length cooled to 0.1 K and coupled to a near-quantum-limited
SQUID amplifier. The red, blue and green lines represent
the thermal, amplifier and total noise, respectively.}
\label{f6-noise}
\end{figure}

Figure~\ref{f6-noise} shows the instrument noise spectral density for SOGRO with $l = 5$ m, $M = 1$ ton
(each test mass), and $Q = 5 \times 10^8$ cooled to 0.1 K and coupled to a dc SQUID with $2\hbar$ noise. The
expected sensitivity of the SOGRO comes to within a factor of 2 from that required for $S = 10$.
The same SOGRO with $Q = 10^9$ coupled to a $1\hbar$ SQUID would meet the sensitivity requirement,
provided all the other noise could be reduced to below its intrinsic noise limit.

In principle, the arm-length restriction imposed by eq. (16) can~ be~\!~ overcome~\!~ by~ positioning~ the~ center of~ the~ mini-


\noindent SOGRO at the same $(x, y)$ coordinates as the interferometer test mass.  This could be done by locating the mini-SOGRO above or below the interferometer test mass.  With a SOGRO with a larger arm-length, the required sensitivity could be reached with more modest parameter values. A detailed analysis of this scheme will be published elsewhere.

It is interesting to see how a mini-SOGRO two orders of magnitude less sensitive to GWs
can help ET mitigate the NN by an order of magnitude. This is because a SOGRO with $l = 5$ m
is quite efficient to detect the Rayleigh NN with $\lambda_{\rm R}/2\pi = 56$ m and SOGRO employs a highly
sensitive superconducting displacement sensor. Although achieving a test mass $Q$ of $10^9$ and
reaching the quantum limit for the SQUID noise is very challenging, it is worth investigating the
SOGRO option since it has intrinsic advantages over seismometers in that it detects the NN directly
and can monitor the local gravity gradient environment with high sensitivity.

\section{Correction of wave-front aberrations in interferometric gravitational wave detectors}
\emph{Operation~ and ~sensitivity~ of~ interferometric~ gravitational }

\noindent \emph{waves detectors can be strongly limited by wave-front aberrations in the core optics. These aberrations are due to intrinsic defects in the optics (surface figure errors and refraction index inhomogeneity) and to the rise of thermal effects and can be compensated for by generating proper optical path length corrections. Here we review the adaptive optical system installed in the largest interferometers and give a glance to future applications in third generation detectors such as ET.}

\subsection{Introduction}\label{introduction}

The largest  interferometric Gravitational Wave detectors Virgo ~\cite{Virgo1} and LIGO~\cite{Ligo1}
are power-recycled Michelson interferometers (ITF) with Fabry-Perot arm cavities. They have operated close to
the initial design sensitivity, completing several observational runs~\cite{Nature,Obs1,Obs2,Obs3}. Further data taking runs with the "advanced" configuration of these detectors are planned on short term time-scale. The second generation interferometers Advanced VIRGO~\cite{AdV} and aLIGO~\cite{4.1aLIGO} will see a significant improvement in sensitivity of about one order of magnitude over the whole detection bandwidth (from 10 Hz to 10 kHz), increasing by a factor of a thousand the number of galaxies explored. This will open the era of GWs astrophysics, since several GW signals emitted by strongly
gravitating systems, such as neutron stars or black holes, are expected to be detected at design sensitivity~\cite{Even}.

In order to reach the design sensitivity, these instruments require at the output port (the ITF antisymmetric port) near perfect destructive interference of the two beams
reflected from the arms.
However, the interferometer beams are degraded by the optical defects in the mirrors, which
results in asymmetries and therefore unwanted power at the antisymmetric port of the detector.

\subsection{Sources of wave-front aberrations}\label{WD}

In interferometric detectors, there are two sources of optical defects:

$\bullet$ errors in the mirror fabrication process (also termed ``cold defects''): imperfections that can occur during the production of the substrate, surface polishing and coating;

$\bullet$ self heating (``hot defects''): the coatings and substrates of the optics absorb a tiny fraction of the power stored in the arm cavities.

\subsubsection{Cold defects}\label{CD}

\begin{figure}[H]
\centering
\includegraphics[scale=0.95]{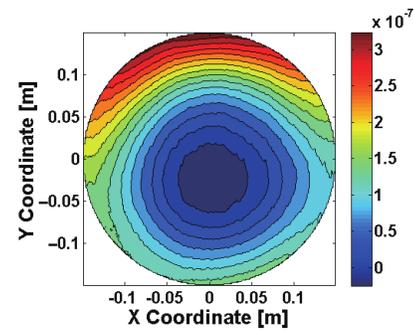}\vspace{-2mm}
\caption{(Color online) Example of a transmission map of a core optic. The color scale is in meters.}\label{fig:sub}
\end{figure}

\begin{figure}[H]
	\centering
	\includegraphics[scale=0.45]{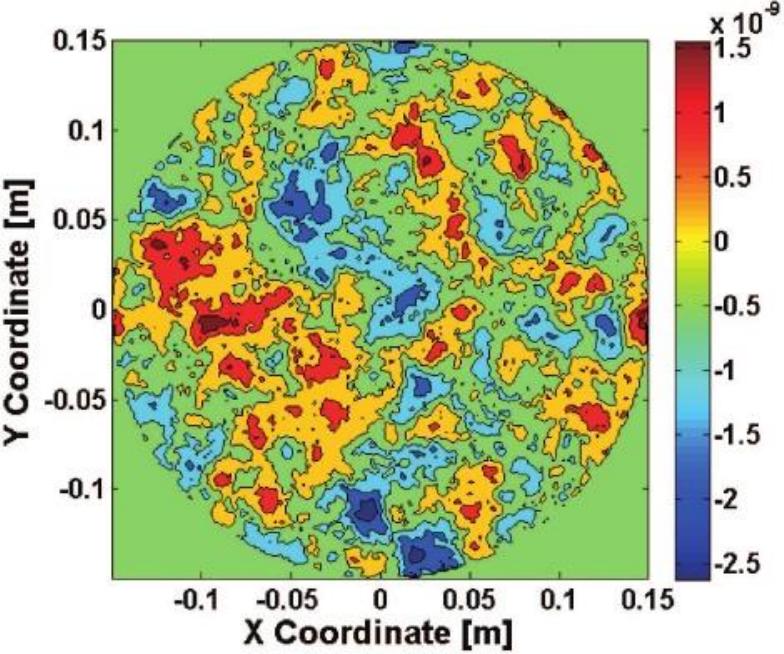}\vspace{-2mm}
	\caption{(Color online) Example of a core optic surface map. The color scale is in meters.}\label{fig:surf}
\end{figure}

A large fraction of optical defects, often without any symmetry, arises from imperfections in the production and polishing of the glass used for the various substrates in the resonant cavities. Surface figure errors on reflective and transmissive surfaces do contribute to the aberrations as well as spatial variations in the index of refraction of the substrates. Figures~\ref{fig:sub} and~\ref{fig:surf} show respectively the optical path length~ increase~ in

\noindent the substrate of a test mass due to the inhomogeneity of the refraction index and to the surface figure error.

\begin{figure}[H]
\centering
\includegraphics[scale=0.6]{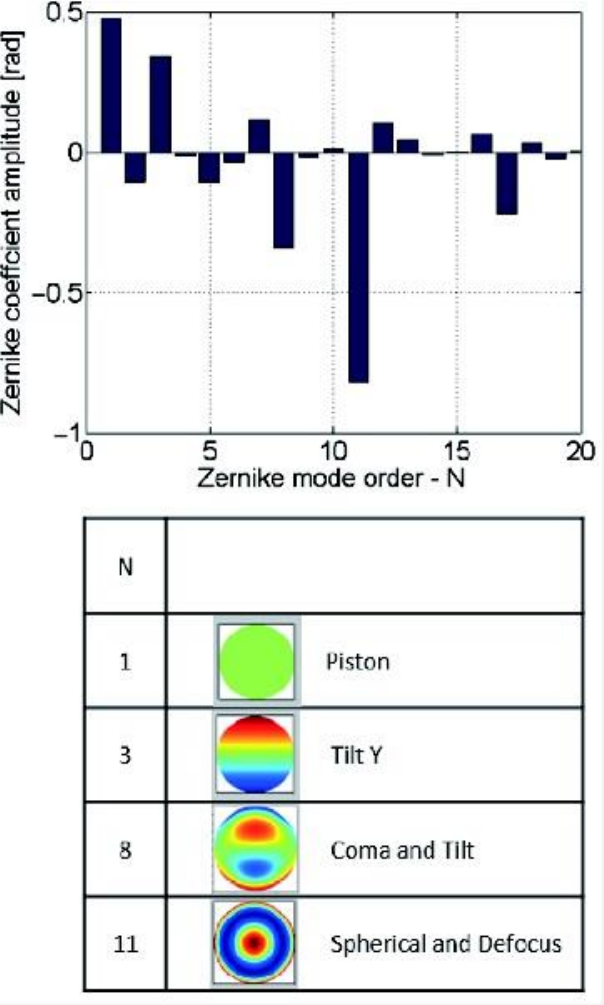}\vspace{-2mm}
	\caption{(Color online) Upper: Zernike coefficient amplitudes of the transmission map in Figure~\ref{fig:sub}. Lower: spatial profile of the modes with the highest amplitudes.}\label{fig:zerni}
\end{figure}

These imperfections may contain both low order (low spatial frequencies) and higher order (high spatial frequencies) aberrations. This is clearly shown when analyzing, for example, the map in Figure~\ref{fig:sub} by evaluating the power spectrum or the Zernike coefficients (see Figure~\ref{fig:zerni}).

\subsubsection{Thermal effects}\label{TL}

The dependence from temperature of the refraction index and the thermal expansion coefficient of optical materials ensure that temperature gradients, induced by the absorption of the Gaussian-profiled probe light, results in nonuniform optical path length distortions.

\begin{figure}[H]
\centering
\includegraphics[scale=0.6]{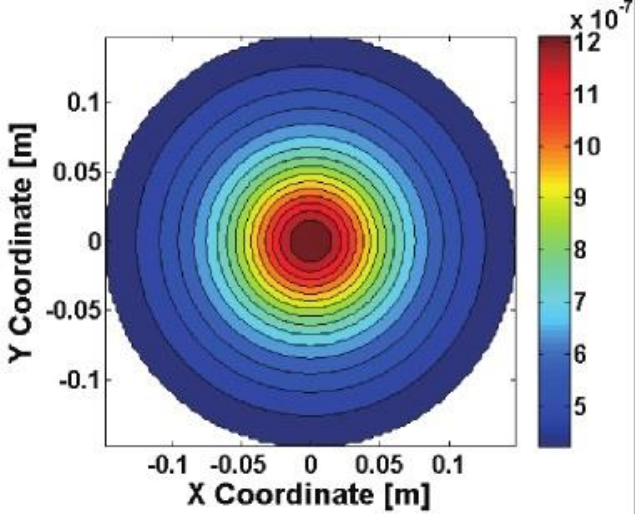}\vspace{-2mm}
\caption{(Color online) Expected wave-front distortion due to thermal lensing at 125 W input power, with 1 ppm absorption, a Fabry-Perot cavity Finesse of 450 and a recycling cavity gain of 36. The color scale is in meters.}\label{fig:therm}
\end{figure}

Thermal effects have already been observed in Virgo~\cite{Virgo1} and LIGO~\cite{Ligo1} and required the installation of Thermal Compensation Systems~\cite{Mau} (TCS). Advanced detectors~\cite{AdV, 4.1aLIGO} will be characterized by a higher circulating power (from 20 kW in the initial interferometers to 700 kW in the second generation detectors) and thermal effects will become even more relevant. The expected wave-front aberration in an advanced detector is shown in Figure~\ref{fig:therm}, for an absorbed power of about 650 mW.

In the test mass, the optical power is absorbed by the substrate and by the high reflectivity coating, the latter contribution being the~ dominant one.~\!~ The absorbed~ power ~is~ con-

\vspace{-5mm}

\noindent verted into heat, producing a temperature gradient inside the substrate.
Two different effects originate from the heating of the test mass:

$\bullet$ nonuniform optical path
	length variations (thermo-optic effect, also termed thermal
	lensing) mainly due to the temperature dependency of the index of refraction.
	
$\bullet$ change of the profile of the high reflective surface, due to thermal expansion (thermo-elastic deformation) in both input and end test masses. It affects the radii of curvature of the test masses.

\subsection{Consequences of wave-front distortions}\label{ConsWD}

Wave-front distortions impact on both the operation and sensitivity of interferometric detectors as they represent a departure from the ideal optical design.

A phase distortion in a resonant cavity acts to scatter power out of the fundamental mode, and thus out of the cavity, and can be viewed as a loss term. In presence of aberrations, which change the cavity mode, the coupling coefficient between the laser TEM$_{00}$ and the cavity TEM$_{00}$ becomes less than one. This leads to a decrease of the cavity gain and thus in the optical power stored in the cavity itself.
For example, in case of the sidebands fields in the recycling cavity, the cavity gain and thus the sidebands power, approximately decrease as~\cite{law}:
\begin{equation}
	G_{\rm rec}=G_{\rm rec}^0\times \frac{1}{\left(1+\frac{r_{\rm pr}\times L}{(1-r_{\rm pr})^2}\right)},
\end{equation}
where $G_{\rm rec}^0$ is the recycling cavity gain in absence of thermal lensing, $r_{\rm pr}$ is the reflectivity of the power recycling mirror and $L$ is the fractional power scattered out from the TEM$_{00}$ mode~\cite{Hello}, termed ``coupling losses'' and defined as:
\begin{equation}
	L=1-A^* A,
\end{equation}
where
\begin{align}
	A=\frac{\langle E_0 | E \rangle }{\langle E_0 | E_0 \rangle } &=\frac{\langle E_0 | {\rm e}^{{\rm i} \frac{2\pi}{\lambda} OPL} |E_0 \rangle }{\langle E_0 | E_0 \rangle } = \nonumber\\
	& = \iint {\rm e}^{{\rm i} \frac{2\pi}{\lambda} OPL(x,y)}|E_0(x,y)|^2 \d x\d y.
	\label{eq:2}
\end{align}
$E_0$ represents the undisturbed cavity field before being subjected to the optical path distortion $OPL(x,y)$ in the recycling cavity and $E$ is the distorted field.
The ultimate consequence is a loss of signal to noise ratio at high frequencies due to the increase of shot noise.

This is true for both the wave-front aberrations in the Fabry-Perot cavities and in the recycling cavity.
In the latter case, another effect takes place: due to aberrations, the interference between the beams reflected by the two arms is less than ideal and unwanted light reaches the antisymmetric port with a consequent increase of noise on the detection photodiodes.

\subsection{Compensation of wave-front distortions}\label{TCS}

\subsubsection{General guidelines}

To correct the optical aberrations, one must somehow induce in the mirrors an optical path length increase equal but opposite to the distortions. A flexible system is needed in order to have the possibility to change the strength and shape of the corrective lens and follow the different interferometer operating conditions, as thermal effects change as a function of the input power.

This can be done by exploiting the thermo-optic effect and depositing heat at specific locations of the mirrors. Not all the techniques are suitable for this goal as the mirrors are the free-fall test masses of our GW detector. A ``touchless" way to heat the mirror is by shining it with a radiation that is completely absorbed (e.g. $\lambda >5\, \upmu$m for fused silica). For example, LIGO and Virgo thermal compensation systems (TCS) used CO$_2$ ($\lambda=10.6\, \upmu$m) lasers, while GEO used a radiating hot element behind a test mass.

\begin{figure}[H]
	\centering
	\includegraphics[scale=0.55]{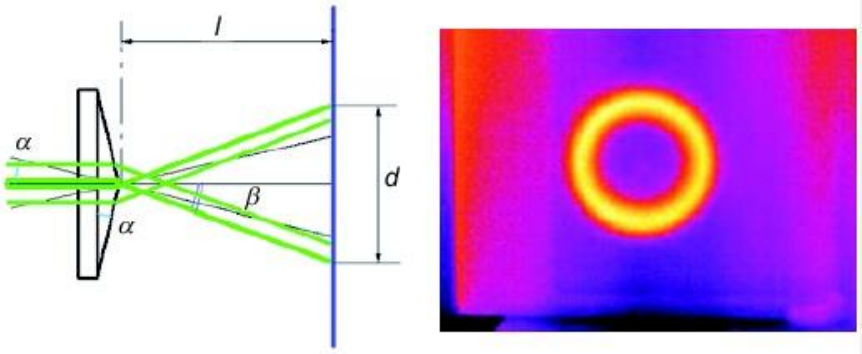}
	\caption{(Color online) Left: working principle of an axicon lens. Right: intensity profile of a laser beam after passing through an axicon.}\label{fig:axic}
\end{figure}

The shape of the compensating pattern depends on the specific distortion that need to be corrected. For instance, in LIGO and Virgo, the only aberration that mattered was thermal lensing in the recycling cavities, that has the shape shown in Figure~\ref{fig:therm}, thus axicon-based optical projectors were used to convert a CO$_2$ laser Gaussian beam into an annular beam~\cite{Mau}. The working principle of an axicon lens and the intensity profile generated by an axicon are shown in Figure~\ref{fig:axic}.

In the most general case, where no particular symmetry is present in the aberrations, Finite Element Modeling can be used to evaluate the corrective heating pattern to be applied to the optics.

\subsubsection{Actuators for TCS}

Different kinds of actuators can be envisaged to correct wave-front distortions. In this section, we will focus on those used in aLIGO and Advanced VIRGO.

(1) Ring heaters.

As stated in sect.~\ref{TL}, to maintain the arm cavity mode structure, it
is necessary to control the radii of curvature of all test masses.

This problem has been already faced in the past: the GEO detector~\cite{GEO} used a ring heater (RH) to change the RoC of one of the two test masses~\cite{GeoRH}. The GEO RH is placed on the back of the mirror, radiatively coupled with the optic.

\begin{figure}[H]
	\centering
	\includegraphics[scale=0.55]{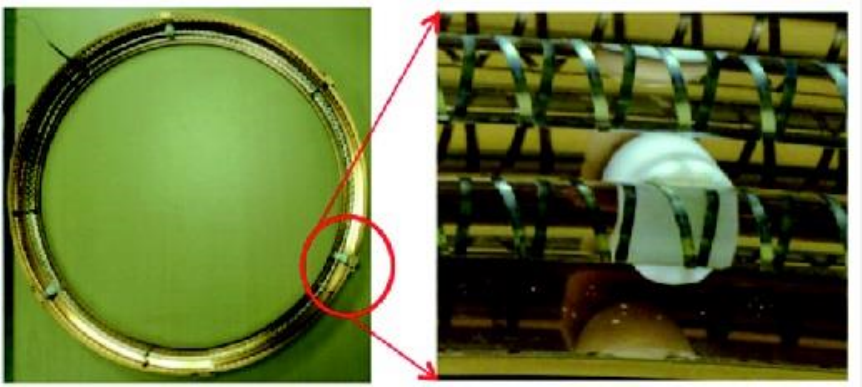}
	\caption{(Color online) Left: Advanced VIRGO ring heater and its radiations shield. Right: detail of the NiCr wire wound on the glass former.}\label{fig:rh}
\end{figure}

Compensation and control of the test mass high reflectivity surfaces will be accomplished in Advanced detectors with the same technique. The TCS baseline design considers four ring heaters, one around each test mass. The input mirror RH also provides limited compensation of thermo-optic effect in the recycling cavities. Unlike the GEO heater, these RHs are equipped with a reflecting shield to maximize the amount of power reaching the test mass~\cite{AdVTCS1,AdVTCS2}. Figure~\ref{fig:rh} shows one of the RHs installed in Advanced VIRGO.

A glass former is used to wind a NiCr wire. The electrical power dissipated in the wire increases the former's temperature, that is radiated on the barrel of the test mass.

(2) CO$_2$ laser projectors.

CO$_2$ laser projectors are used to heat the
optic since the TEM$_{00}$ mode of the laser can be efficiently converted into a
mode with a different intensity distribution by shaping the laser beam. This makes it possible to generate the desired heating profile.

There are several methods to produce the corrective heating pattern, such as binary masks or refractive optical elements. Another method used to generate the heating patterns (especially to correct for non-symmetric aberrations) is based on a CO$_2$ laser scanning system. This technique, developed at MIT~\cite{law}, comprises a pair of galvanometer mirrors, to move the laser beam on the surface of the optic, and an acousto-optic modulator to modify the power content of the beam.

Due to noise requirements~\cite{TCSnoise}, in advanced detectors it is no longer possible to heat directly the test masses. Thus, an additional optic is placed in the recycling cavities, named Compensation Plate (CP), shined with the corrective beams to create within the CP itself the required thermal
lens and, thus, to heal the optical aberrations in the recycling cavities.

This actuation scheme (RHs and compensation plates) reduces the coupling between the
two degrees of freedom (wavefront aberrations in the recycling cavities and changes
of the radius of curvature in the Fabry-Perot cavities) and allows for a nearly diagonal
control matrix.

\begin{figure}[H]
	\centering
	\includegraphics[scale=1]{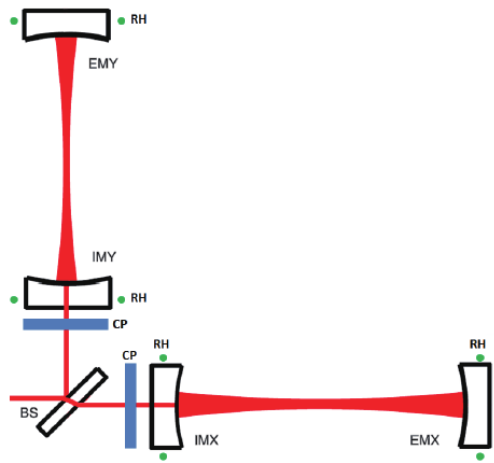}
	\caption{(Color online) Actuation scheme of the TCS in advanced detectors: blue rectangles represent the CPs (heated by the CO$_2$ lasers) while the green dots around the test masses are the ring heaters.}\label{fig:advtcs}
\end{figure}

Following the above considerations, the conceptual actuation scheme of the compensation system designed for advanced detectors~\cite{AdVTCS1, AdVTCS2} is shown in Figure~\ref{fig:advtcs}.

\subsubsection{Sensing wave-front distortions}

The aberrations in the interferometer cavity optics can be sensed by several complementary techniques. For instance, each optic can be independently monitored with dedicated wave-front sensors, while the intensity distribution and phase of the fields in the recycling cavity (carrier and sidebands) can be measured by phase cameras~\cite{phase1}. Moreover, the amplitude of the optical path length increase appears in some ITF channels, such as the power stored in the radio frequency sidebands: these are scalar quantities that can give a measurement of the amount of power scattered into higher order modes.
The TCS control loop has to be designed to use a blend of the signals from the different sensors.

Phase cameras are high-resolution wave-front sensors that measure the complete spatial profile of
any frequency component of a laser field containing multiple frequencies. The basic principle behind this wave-front sensing technique
is to measure the beating between the field under test and a reference field that is spatially overlapped
with it. The high spatial resolution is achieved
using a reference field with high modal purity and a high spatial resolution scan.
Frequency discrimination is
realized by heterodyne detection, which is used to measure
the beat note between the reference field and the
frequency component of interest of the test field.

In interferometric GW detectors, the carrier is resonant in the arm
cavities, which are very
effective spatial filters, while the RF sideband fields
resonate only in the power-recycling cavity and
experience less spatial filtering. Thus, the
RF sidebands are more sensitive to wave-front distortions in the power-recycling
cavity than the carrier field. Consequently, the spatial
modes of the carrier and RF sideband fields exiting
the interferometer may be quite different.
It is, therefore, desirable to measure the spatial modes of
the RF sideband and carrier fields, in order to have a picture of the wave-front distortions in the recycling cavity.

Hartmann Wave-front Sensors (HWS)~\cite{HWS} can give a direct measurement of wave-front distortions through an auxiliary probe beam (at a different wavelength than the ITF beam) that interrogates the optic under test (either in transmission or reflection) and a sensor that measures the wave-front distortions accumulated on the auxiliary beam itself.

\begin{figure}[H]
	\centering
	\includegraphics[scale=0.55]{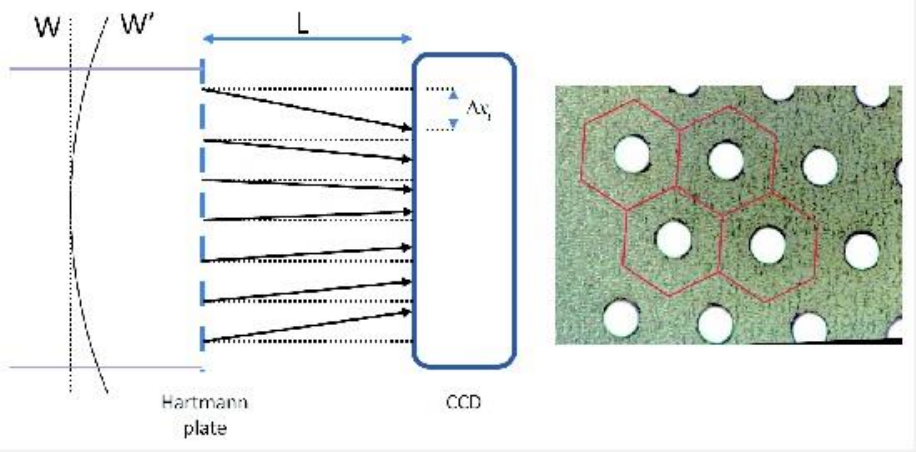}
	\caption{(Color online) Left: working principle of a Hartmann sensor. Right: zoomed view of a Hartmann plate}\label{fig:hws}
\end{figure}

An aberrated wave-front $W$' is incident on a Hartmann plate (essentially a plate containing a series of apertures, see Figure~\ref{fig:hws}).
The resulting rays propagate a distance $L$, normal to the local wave-front gradient, and are incident on a \linebreak
\vspace{-4mm}

\noindent CCD. The new spot position is measured and compared to a reference spot position, determined using a non-aberrated wave-front $W$. The  set of all spot displacements makes the gradient field of the wave-front, which can be
integrated to obtain the incident wave-front.

The key to the wave-front measurement is to relay image the measured optical surface onto the Hartmann plate. If $M$ is the magnification of the optical system between the measured optical surface and the wave-front sensor surface, then the wave-front gradient
\begin{equation}
\left(\frac{\partial \Delta W}{\partial x},\frac{\partial \Delta W}{\partial y}\right)
\end{equation}
at coordinate $(x, y)$ on the optic will result in a measurement of the gradient
\begin{equation}
\left(M \frac{\partial \Delta W}{\partial x},M \frac{\partial \Delta W}{\partial y}\right)
\end{equation}
at sensor coordinates $(x/M, y/M)$. In other words, the wave-front distortions are directly mapped between the optic under test and the sensor.

The Hartmann sensor selected for Advanced VIRGO and aLIGO is that already developed and characterized on test bench experiments and in the Gingin High Optical Power Test Facility for the measurement of wave-front distortion~\cite{brooks_hws}.

This sensor has been demonstrated to have a shot-to-shot reproducibility of $\lambda$/1450 at 820 nm,  which can be improved to $\lambda$/15500 with averaging, and with an  overall accuracy of $\lambda$/6800~\cite{munch}.

\subsection{Beyond advanced detectors}

Different studies are investigating possible upgrades to further increase the sensitivity of future interferometers: instruments keeping the existing infrastructures~\cite{Rana2012} or completely new facilities~\cite{ET}.

In cryogenic detectors, with silicon test masses, it is likely that thermal lensing will become negligible: the high thermal conductivity and the low thermo-optic coefficient will make any optical path length increase negligibly small, even in presence of asymmetric heating~\cite{ET}.

``Xilophone'' configurations have also been proposed~\cite{Hild2011}, to capture the best features and the low and high frequency instruments. In this configuration, the low frequency-low temperature detector shares the site with the high frequency-high temperature interferometer. In the latter, the use of higher order Laguerre-Gauss modes (LG$_{33}$) has been considered~\cite{Sorazu2013}.
The high frequency instrument would be affected by both thermal effects and by the low spatial
frequency mirror figure errors. In fact, it has been shown~\cite{Sorazu2013,Hong2011,Bond2011} that polishing techniques need to improve by at least a factor of ten to reach the required purity of the Fabry-Perot cavity
mode.

Thus, next generation adaptive optical systems will need to reproduce corrective heating patterns with increasing complexity.
Different methods are being investigated, such as arrays of radiative heating sources~\cite{Day2013} or Micro-Electro-Mechanical-System (MEMS) deformable mirrors~\cite{Rocchi2014}, which imprint a phase modulation to the wave-front of the incoming Gaussian
beam that converts the beam's intensity distribution into the desired one holographically.

\subsection{Summary}\label{sec_fin}

Wave-front distortions are an unavoidable annoying presence in interferometric GW detectors: these can be due to the mirror fabrication process or to the absorption by the mirrors of a tiny fraction of the power stored in the ITF. Since aberrations represent a departure from the ideal optical design, the sensitivity and the robustness of the detectors are strongly affected.

Optical simulations and finite element modeling become essential tools to derive the needed heating patterns to be shined on the optics and to evaluate their efficiency.

Thermal compensation systems have been put in operation in first generation interferometers; upgraded configurations have been
installed in advanced detectors to correct wave-front distortions in all core optics, using thermal actuators (Far-Infra-Red lasers and ring heaters) and wave-front sensors (Hartmann sensors and phase cameras).

Further improvements are being investigated for possible implementation in third generation interferometric GW detectors.

\section{Three mode parametric instability and their control for advanced gravitational wave detectors}
\emph{Three mode parametric instability in advanced laser interferometer GW detectors is a technical problem that needs to be controlled in all future high optical power GW detectors. The phenomenon causes ultrasonic acoustic modes in test masses to be driven by high power laser beams so that they ring up exponentially, eventually causing the interferometer locking to fail. This paper reviews the physics of three mode interactions, and reports on the phenomenon of parametric instability observed in aLIGO. Methods for control of instabilities are discussed.}
\subsection{Introduction}\label{introduction}

\begin{figure}[H]
\centering
\includegraphics[scale=1]{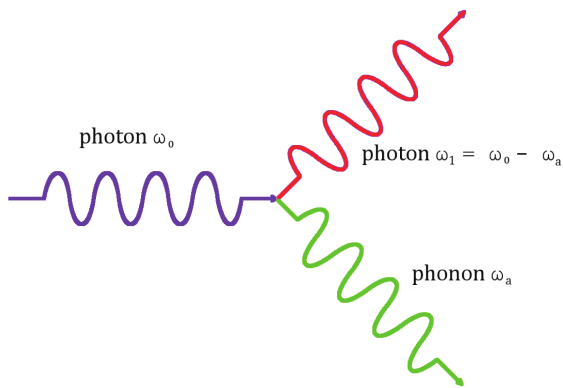}\vspace{-2mm}
\caption{(Color online) The quantum picture of the three-mode parametric interaction: the interaction can be treated as a simple photon-phonon scattering process in which a carrier photon scatters from the acoustic phonon on the mirror surface to create a transverse optical mode photon. The acoustic mode frequency is exactly equal to the difference between the two optical frequencies.} 
\label{fig:1a}
\end{figure}

\begin{figure}[H]
\centering
\includegraphics[scale=0.5]{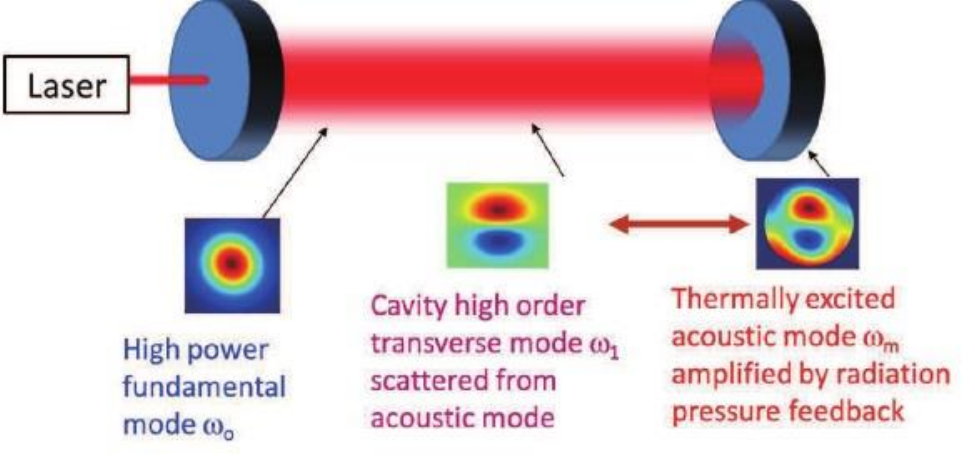}\vspace{-2mm}
\caption{(Color online) Classical picture of a cavity three-mode interaction: the main high power cavity optical mode beats with a transverse mode generated by light scattered from a mirror acoustic mode. The beat frequency causes a time varying radiation pressure force in phase with the acoustic mode. This drives the mirror and excites the acoustic mode as long as the transverse optical and acoustic mode shapes are similar.} 
\label{fig:1b}
\end{figure}

High power operation of advanced GW detectors such as aLIGO and Advanced VIRGO will be dependent upon finding a solution to the problem of the radiation pressure induced parametric instability (PI).  This instability was predicted in 2001 by Braginsky et al. \cite{Braginsky}, who postulated that the very large contained power of the proposed advanced detectors will create large radiation pressure effects.  The predicted instability is caused by a three mode interaction between two optical modes in a cavity and an acoustic vibrational mode in the mirror (test mass) of the cavity. Essentially the intense laser light can scatter inelastically from macroscopic acoustic thermal motion of a mirror such that the photon energy is divided between a lower frequency transverse optical photon and an acoustic phonon in the mirror, as illustrated in Figure~\ref{fig:1a}. In other words, the radiation pressure of the beat note between the fundamental and the scattered high order optical mode (HOOM) in the cavity drives the acoustic mode (Figure~\ref{fig:1b}).


The phenomenon is similar to Brilliouin scattering, but is macroscopic, and occurs in the ultrasonic regime $\sim$ 10--200 kHz. If the acoustic power injected by this mechanism exceeds the acoustic losses of the mirror, the mirror acoustic amplitude will grow exponentially, steadily increasing over seconds or minutes, until a very large amplitude (of say 1 nm) causes saturation of amplifiers and failure of the instrument's control system.
The instability can be thought of as a feedback loop \cite{EvensPIgeneral} with some open loop gain R which we call here the parametric gain (PG):
\begin{equation}
R = \frac{ 8P}{McL}\frac{Q_{\rm m}}{\omega_{\rm m}^2}Q_{\rm o}\Delta_{\rm s}\Delta_\omega \label{parametricGain},
\end{equation}
where $Q_{\rm m}$ is the quality factor ($Q$) of the mechanical mode, $P$ is the contained power in the fundamental optical mode of the cavity, $M$ is the effective mass of the acoustic mode, $c$ is the speed of light, $L$ is the length of the cavity,  $\omega_{\rm m}$ is the mechanical mode frequency, $\Delta_\omega$ is the frequency overlap term and $\Delta_{\rm s}$ is the spatial overlap term described later.  The factor $\Delta_\omega$ is given by
\begin{equation}
\Delta_\omega=\frac{1}{1+ \omega_\Delta/\delta_{\rm o}} \label{eq:deltaOmega}.
\end{equation}
Here $\delta$ is the half linewidth of the HOOM,  $\omega_\Delta=\omega_0-\omega_1-\omega_m$, with $\omega_0$, $\omega_1$, $\omega_m$ being the frequencies of the $TEM_{00}$ fundamental optical mode, the HOOM $TEM_{mn}$ and the acoustic mode respectively.

In 2005 the group in University of Western Australia (UWA) undertook a detailed 3D simulation of parametric instability \cite{ZhaoPI}, predicting that detectors like the planned aLIGO GW detector would indeed experience a three-mode opto-acoustic parametric instability, involving tens of acoustic modes across the four main interferometer test masses.  At the Gingin facility  the UWA team (with the support of the Australian Consortium for Interferometric Gravitational Astronomy) developed 80 m long cavities with kg-scale test masses and increased optical power especially designed to investigate this phenomenon.  In late 2014, spontaneous parametric instability in a large suspended optical cavity was first observed in the Gingin 80 m cavity \cite{ginginPI}.

\begin{figure}[H]
\centering
\includegraphics[scale=0.52]{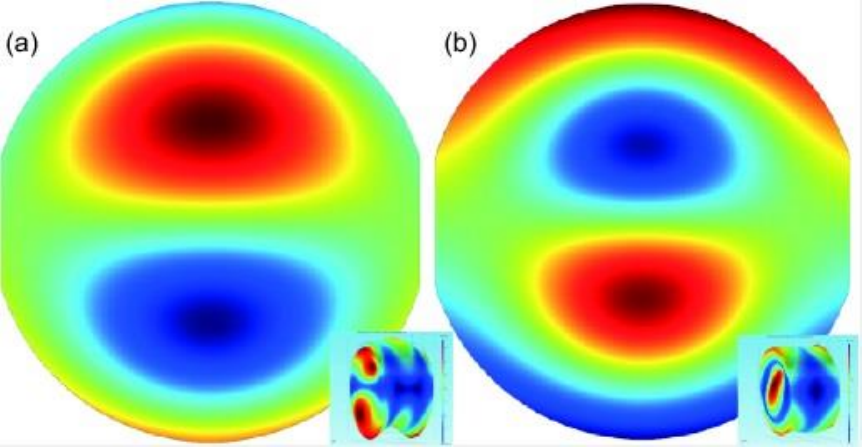} \vspace{-1mm}
\caption{(Color online) The simulated (COMSOL) tangential surface deformation of the two acoustic modes responsible for parametric instability in aLIGO. (a) is the ~15.00 kHz acoustic mode and (b) is the ~15.53 kHz acoustic mode, insets are total surface deformation in 3D.} \label{AcousticModes1}
\end{figure}

It takes the extreme technology of long baseline laser interferometers to enter the regime where instability can occur: 40 kg scale mirrors with ultralow acoustic losses, 4 km long optical cavities, and very high optical power $\sim$ hundreds of kilowatts. In November 2014 parametric instability became a reality for operations of the aLIGO detectors. Shortly after the observation at the Gingin facility, instability was observed \cite{EvansPIobs} in the aLIGO Livingston facility with only $10\%$ of the designed full power during commissioning phase of the detector. As expected the instability appeared as an exponential ring up of the acoustic modes of test masses. The interferometer lost lock when the acoustic mode of Figure~\ref{AcousticModes1}(b) amplitude was $\sim4.5\pm1$ orders of magnitude above  \linebreak
\vspace{-4mm}

\noindent it's thermally excited level.  PGs up to 2.2 were observed with an operating injected power of $25$ W.  It is predicted that if the interferometer was tuned so that $\omega_\Delta=0$, the PG would be $3\pm2$  for the 15004 Hz mode and $7\pm3$ for the 15536 Hz mode.  The larger of these two PGs would result in a ring-up time costant of $\sim$2 min and an interferometer operating time of $\sim$20 min at 25 W assuming acoustic modes start in their thermally excited state.  This paper presents experimental results that characterise the state of the interferometer with regards to parametric instability.  Tolerances on estimates of the quality factor of the mechanical modes and the frequncy spacing between the fundamental and high order optical modes must be improved before firm predictions about high power operations of these detectors.

This section is organized as follows: Sect.~\ref{control} summarise the control strategies that have be explored over the years.  In sect.~\ref{acc}, we explore the behavior of the acoustic modes in the interferometer test masses. In sect.~\ref{optical_modes}, we present the theoretical and measured behavior of optical modes in aLIGO arm cavities. Then in sect.~\ref{Overlap} the spatial overlap is considered, these results are combined into a predicted PG with experimental verification in sect.~\ref{Parametric_Gain}. Finally the results are summarized and prospects for the future are considered in sect.~\ref{sec_fin}.

\subsection{Control of parametric instability}\label{control}

From eq. ~(\ref{parametricGain}) it can be seen that many of the design requirements for advanced interferometers, such as high power for lower shot noise, high $Q$ materials for low thermal noise etc increase PGs.  It also indicates a set of control strategies:

$\bullet$ Reduce the power

$\bullet$ Change the frequency overlap

$\bullet$ Change the spatial overlap

$\bullet$ Reduce the mechanical mode $Q$ factor

$\bullet$ Reduce the finesse of the fundamental or higher order optical mode

$\bullet$ Suppress the HOOM

$\bullet$ Suppress the mechanical mode amplitude.

 Several parametric instability control/suppression schemes were proposed:

 (a) Ring damper method\cite{ringdamper}.  Because the threshold for instability is proportional to the acoustic $Q$-factor $Q_{\rm m}$ of the test masses, acoustic damping is a logical approach. The ring damper involves coating a thin layer of lossy material at the circumference of the test mass.  This will reduce the $Q$ factor of the acoustic mode.  However, analysis shown that this may introduce unacceptable thermal noise level increase.

 (b) Another damper method is the tuned resonance dampers \cite{resonantdamper}, which consists of several small resistively-shunted piezoelectric dampers attached to the test mass.  Preliminary tests showed that this method has potential \cite{resonantdamper2}, but the high mechanical loss of the piezo-electric and epoxy materials used in the tests would cause an unacceptable increase in aLIGO thermal noise \cite{damperGras}.

 (c) Optical feedback suppression \cite{optPI} were demonstrated, which would suppress instability by suppressing the build up of the transverse optical mode associated with the instability. This method involves detecting the onset of instability, generating an interference beam and injecting it into the optical cavities. This sounds simple, but is complicated by the need to generate arbitrary transverse modes of precise frequency and to minimise the noise injection.

 (d) Many low gain instabilities may be controllable by direct electrostatic feedback \cite{Electrostatic} to the test masses using electrostatic drive plates already installed in aLIGO.  Its operation is dependent on the magnitude of the overlap integral achievable between the actuator and the thousands of potentially unstable acoustic modes. Because the overlap integral depends strongly on relative locations of the electrostatic combs it is difficult to be certain that it will have sufficient overlap for all potentially unstable modes. Such experiments are already underway at LIGO.

 (e) Thermal tuning \cite{ZhaoPI,JeromeThermal} for parametric instability control was proposed and extensively investigated \cite{thermaltuning} at the 80 m high power cavity \cite{gingin} of the Gingin facility.  This method uses heating to change the mirror radius of curvature (RoC) so as to change the frequency spacing of the optical modes of a cavity, and thus frequency overlap $\Delta_{\rm s}$ in eq.~(\ref{parametricGain}). The method is now fully confirmed and has been used frequently at the Gingin facility to tune three mode interactions. At the LIGO detectors, ring heaters near the test masses are used to tune the radius of curvature of the test mass mirrors instead of direct surface heating. However in a long baseline interferometer the mode density is so high that thermal tuning generally tends to transfer the instability from one mode to another. Thus a combination of  control techniques are required.

Exact predictions for PI have been difficult, predominantly due to the sensitivity of the HOOM frequency to mirror RoC, but also due to unmeasured parameters like the $Q$ factors of the acoustic modes.  Previously statistical approaches have been used to predict the severity of PI by the metric of number of unstable acoustic modes \cite{GrasDetail}.
To some extent parametric instability control is hampered by lack of knowledge of which acoustic and optical modes are likely to be unstable, since this depends very strongly on unpredictable details such as thermal deformations and alignment variations. It would be very useful to have a means of diagnosing and predicting parametric instability before it occurs such as the low power PG estimates made in ref. \cite{CBlair13}.

As mentioned before, parametric instability occurs when the 3-mode parametric interaction gain $R>1$.  However, even when there is no instability, there would be numerous mostly low gain acoustic modes that are detectable through 3-mode interactions. It was shown \cite{3MImonitor} that the three mode interactions normally present in large scale GW detector arm cavities can be used to create a precision monitoring tool which is sensitive to small variations in mirror radii of curvature and spot positions. With the help of thermal tuning of the test masses, it is possible to identify the potential unstable acoustic modes of each test mass.  Monitoring these modes through 3-mode interaction would provide indication for instability for implementation of control schemes.

\subsection{Acoustic modes of the test masses}\label{acc}

The eigen-frequencies for rotationally symmetric eigen-modes of a cylindrical body were described by Chree  \cite{Chree} in 1886:
\begin{equation}
	\frac{k}{2\pi} = \frac{p}{2\pi}(2n/\rho )^{\frac{1}{2}},
\end{equation}
where $k/2\pi$ is the resonant frequency, $p$ is a spatial parameter that has certain allowed values, $\rho$ is the density and  $n = E/(\sigma+1)$ with $E$ the Young's modulus and $\sigma$ the poison ratio.

For aLIGO and Virgo detectors, the chosen material of the mirror test mass is fused silica.  Fused silica has very high acoustic $Q$-factor.  It also has an interesting property that the Young's modulus, rather than continually decreasing with temperature, has an inversion from $-$200 to ~1000 degrees \cite{SpinnerT_Y}.  Around 17$^{\circ}$C there is therefore a relation $\frac{\; \partial k}{k \partial T} = \frac{\; \partial n}{2n \partial T} \approx 7.7\times10^{-5}$ Hz/deg between acoustic mode frequency and temperature that turns out to be very useful in experimental investigations.

The acoustic modes of test masses can be monitored in various outputs of the interferometer.  When these acoustic modes have large amplitudes they couple into a swath of output channels, most notably they are down-converted through aliasing into the detection band of the main interferometer sensing channel which is how PI was first detected \cite{EvansPIobs} with the acoustic mode of the Y end test mass (ETMY) at LIGO Linvingston at 15538 Hz.  Since this observation parametric instability has been observed with another acoustic mode at 15004 Hz in ETMX. Both these instabilities were due to the 3$^{rd}$ order optical mode around 15 kHz.  An investigation comparing measured acoustic mode frequencies and COMSOL finite element simulation leads us to believe the  surface profiles of these modes are those depicted in Figure~\ref{AcousticModes1}.

\begin{figure}[H]
\centering
\includegraphics[scale=0.31]{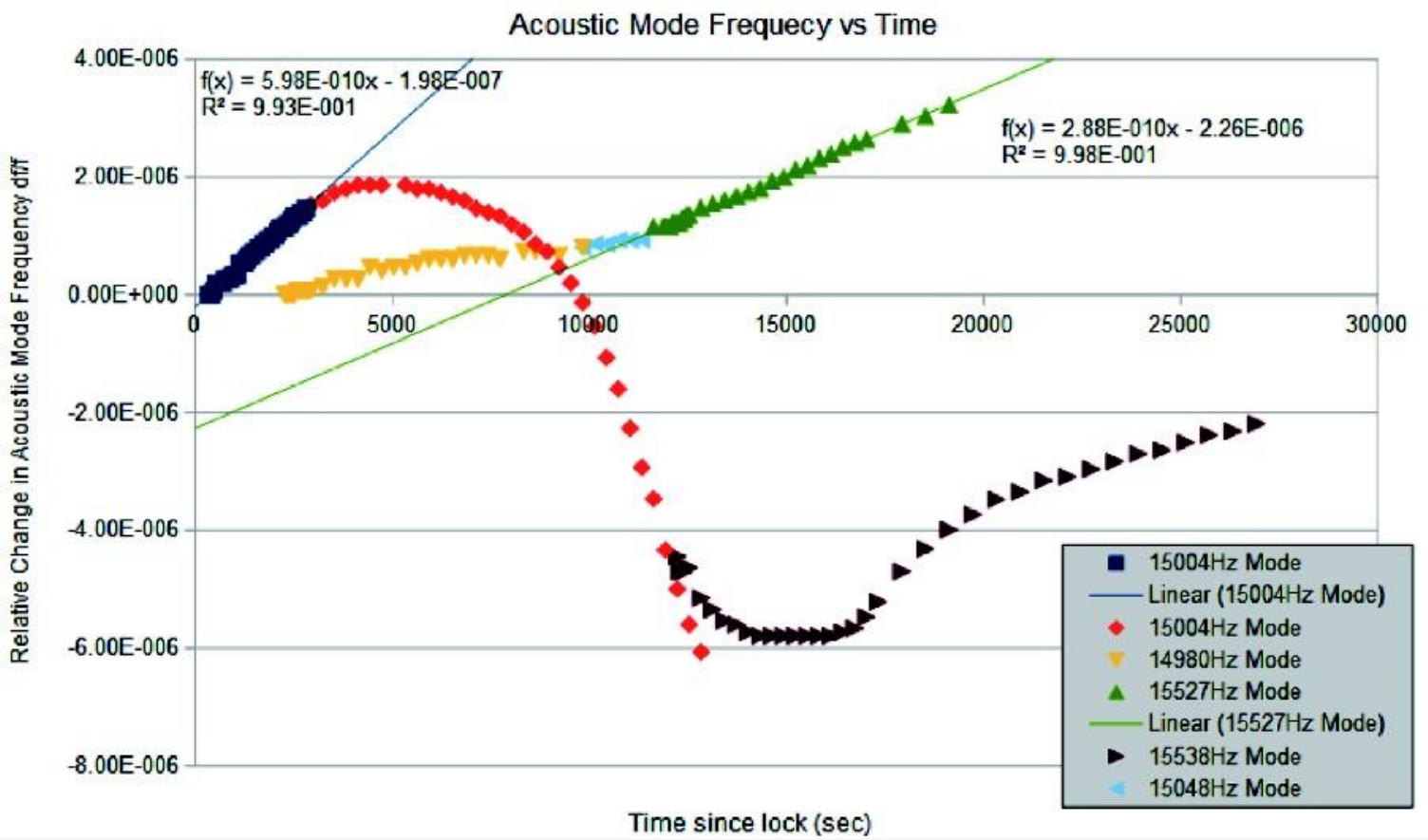}\vspace{-2mm}
\caption{(Color online) An example measurement of the response of various acoustic mode frequencies (measured in transmission of the arm cavities through the three mode interaction) to steps in ring heater power.  The modes can be seen to fall into two groups those that vary with the ring heater and self heating and those that vary through solely self heating from the contained laser power in the interferometer}
\label{actemp}
\end{figure}

 The test mass associated with each acoustic mode can be identified by correlating the observed frequency shift to the temperature of the optics. This was done by either modulating or changing the ring heater power, or by correlating with the ambient temperature. In the example below the ETM ring heaters were stepped three times: up at the beginning, down at $\sim3000$ s and up again at $\sim11000$ s. The observed 5 acoustic modes frequency vary in two groups as shown in Figure~\ref{actemp}:  those affected by the ring heater power change (modes 15004 Hz and 15538 Hz) and those unaffected by the ring heater but only vary slowly through heating from the main cavity laser (modes 14980 Hz, 15058 Hz and 15527 Hz). Thus it can be deduced that modes 15004 Hz and 15538 Hz are the acoustic modes of the ETM being heated by the ring heater.

\begin{figure}[H]
\centering
\includegraphics[scale=0.45]{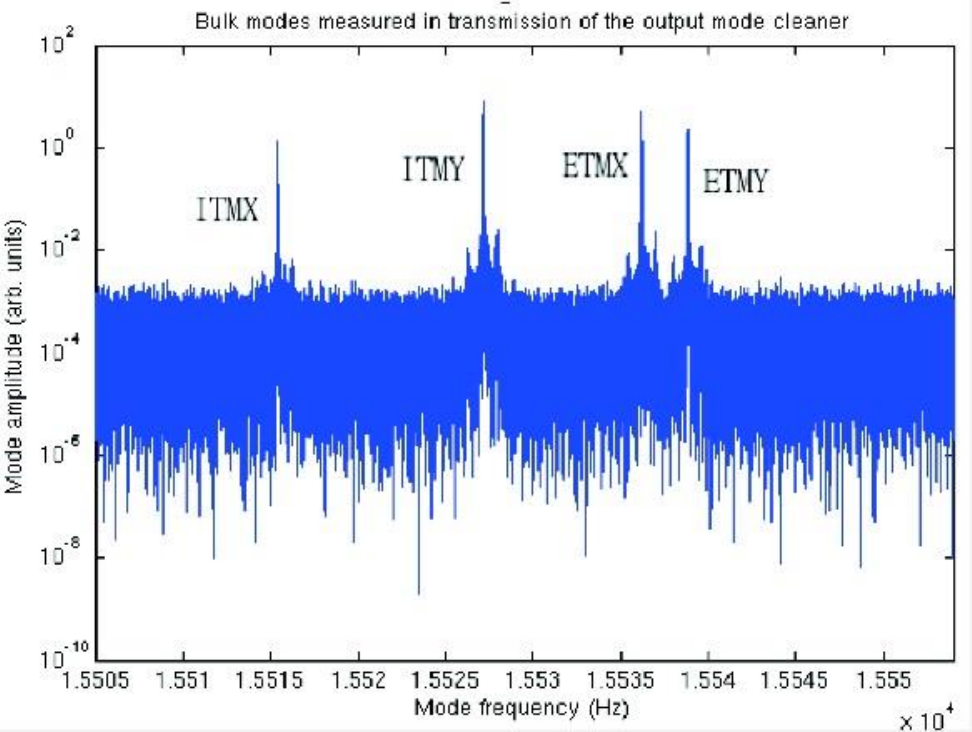}\vspace{-2mm}
\caption{(Color online) An example measurement of the OMC transmission depicting a group of 4 acoustic modes associated with the 4 test masses, associated with the 15.53 kHz acoustic mode depicted in Figure~\ref{AcousticModes1}(b).}
\label{acOMC}
\end{figure}

From these investigations we were able to identify most of the simulated acoustic modes and determine the test mass associated with each.  The two acoustic modes responsible for instability are therefore the ETMY mode of the form in Figure~\ref{AcousticModes1}(a) at 15004 Hz and an ETMX mode of the form in Figure~\ref{AcousticModes1}(b) at 15538 Hz, with the arm being identified by the relative amplitudes of the signals in the arm transmission signals.  The $Q$ of these modes can be estimated from the line-width.  The output of the output mode cleaner (OMC) is remarkably sensitive to the acoustic modes as shown in Figure~\ref{acOMC}.

From Figure~\ref{acOMC} we see that there is ample signal to noise ratio to measure the acoustic mode line-width.  However there is a problem in measuring line-widths of the order $0.001$ Hz. Long stretches of data are required and as can be seen in Figure~\ref{actemp} the acoustic mode thermal frequency shifting would result in smearing of the acoustic mode peak on these time scales.  It was found that the minimum line-width was recorded with $\sim1800$ s of data.  These line-width measurements give a lower bound on the $Q$-factor of $6.8 \pm 1 \times 10^6$ for the ETMX 15004 Hz mode and $7.1 \pm 1 \times 10^6$  for the ETMY 15538 Hz mode. The $Q$ estimate is re-examined through the parametric interactions and is presented in sect. ~\ref{Parametric_Gain}.
The other piece of information that can be gained from Figure~\ref{acOMC} \linebreak
\vspace{-4mm}

\noindent is a calibration of the OMC DC photodiode.  By assuming that the acoustic mode is thermally excited at the beginning of the lock stretch, the equipartition theorem can be applied as in ref.~\cite{DoolinOptCanti} to estimate the calibration factor:

\begin{equation}
	\alpha^2 = \frac{4k_{\rm  B}TQ_{\rm m}}{m_{\rm eff}\omega_0^3},
\end{equation}
where $\alpha$ is the calibration factor in unit of cts/m, $k_{\rm B}$ the Boltzmann constant and $m_{\rm eff}$ the effective mass of the mode.  The effective mass of the 15.54 kHz acoustic modes can be estimated as 9.047 kg using COMSOL and the method described by Hauer et al. \cite{Hauer}.  The calculated calibration factor is $13\times10^{12}$ cts/m for the peak surface displacement on the mirrors surface.

\subsection{Optical mode spacing}\label{optical_modes}

For the 4 km long arm cavities of aLIGO the free spectral range $F_{\rm fsr}= c/2L$ is 37500 Hz.  The line-width of the cavities have been measured several times \cite{Martinov} and is significantly position dependent \cite{Tsukada}.  Using a nominal central alignment the finesse $\mathcal{F} = 417\pm3$ and thus the line-width $\delta f = f/Q = F_{\rm fsr}/\mathcal{F} =89.9\pm0.6$ Hz.  For the purposes of this paper the HOOM line-width is assumed to be the same of the fundamental mode line-width.
The supported HOOM are assumed to be transverse electromagnetic (TEM) modes in the Hermite Gaussian basis ($l,m,n$). They have a frequency that depends on the g-factor of the cavity with the following relation:
\begin{equation}
f_{mn} = F_{\rm fsr} (m+n+1) /\pi \cos^{-1}(\sqrt{g_1 g_2 } )\label{eq:hoomSpa},
\end{equation}
where $g_{1,2} = (1-L/R_{1,2})$ and $R_{1,2}$ are the RoC  of the mirrors.

Using values measured pre-installation$^{2)}$ \footnote{2) https://galaxy.ligo.caltech.edu/optics/, accessed Aug 2015} the theoretical mode spacing for the X and Y arms are 14944 Hz and 15137 Hz respectively.  However the HOOM frequency is very sensitive to the RoC of the optics which itself is sensitive to the thermal state of the optics.  The large laser power build-up in the cavity directly heats the mirrors through coating absorption and this results in a thermal deformation \cite{ZhaoThermLens}. Such a deformation was described analytically by Hello et al. \cite{4.1Hello}.  The finite element analysis package COMSOL has also been used to simulate the thermal transient of the test mass.  It can be seen from Figures~23 and 24 that there are large discrepancies between simulation and the measured thermal state of the optics as can be seen from the measurements of the acoustic mode frequency and measurements of the LG20 mode made during early testing on the interferometer \cite{Brooks}.
We would like to determine the HOOM spacing in situ.  Although the steady state thermal response may not be well understood, the first two hours are almost linear.  Assuming that the long time constant variability can be attributed to factors external to the optics \cite{Yu} and we can trust the model in this early period, \linebreak
\vspace{-2mm}

\begin{figure}[H]
\centering
\includegraphics[scale=0.45]{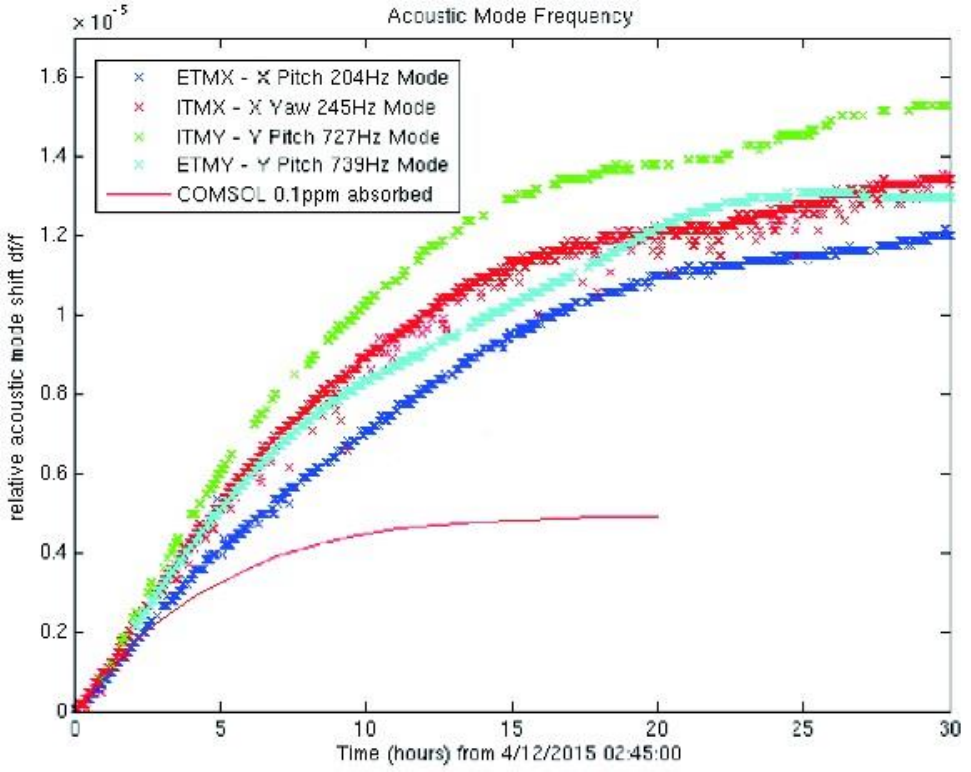}\vspace{-2mm}
\caption{(Color online) The acoustic mode frequency shift from the laser absorption of the coating when the cavity contained power was stepped to ~100 kW with simple COMSOL simulation at 100 ppb.}  \label{OpticalModeSp}
\end{figure}

\begin{figure}[H]
\centering
\includegraphics[scale=0.48]{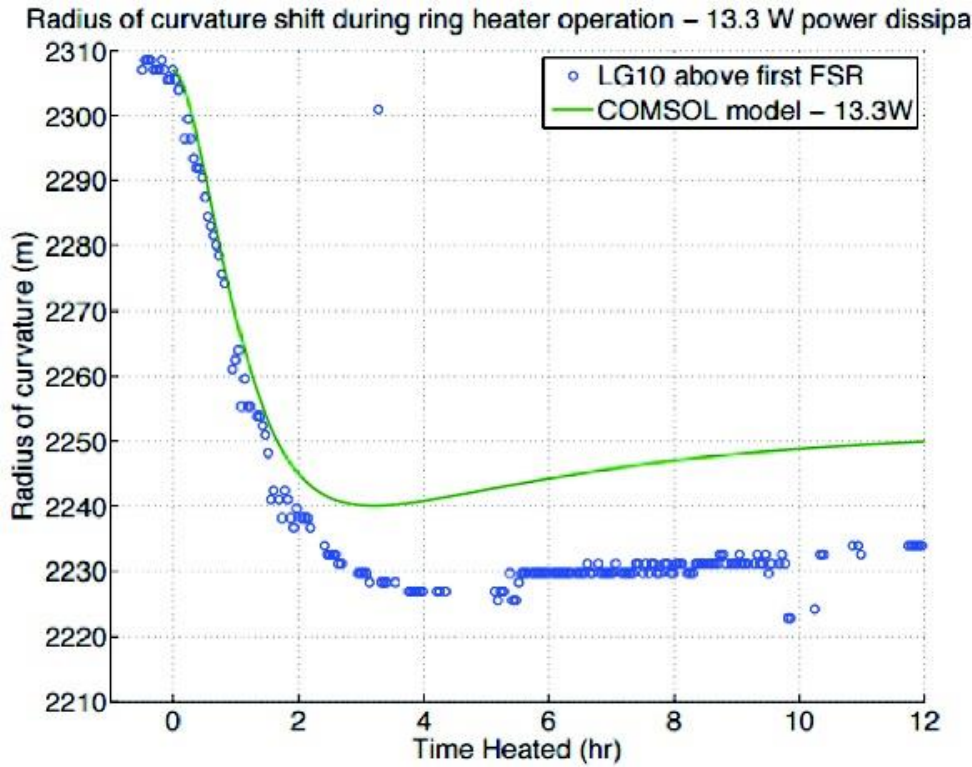}\vspace{-2mm}
\caption{(Color online) Third order optical mode spacing in a single LIGO arm cavity, simulation and experimental results (copied from Brooks\cite{Brooks}).}  \label{OpticalModeSp}
\end{figure}

\noindent then the response of the RoC to a change in the ring heater power has a rate of $(3.0 \pm 0.2)$ mW$^{-1}$ h$^{-1}$.  The ring heater can therefore provide an approximately linear HOOM frequency sweep between 15 min to 90 min after a step in ring heater power.  By estimating the PG of acoustic mode over the duration of the sweep the data can be fitted to the cavity resonance to estimate the HOOM frequency at the original ring heater power.

\begin{figure}[H]
\centering
\includegraphics[scale=0.5]{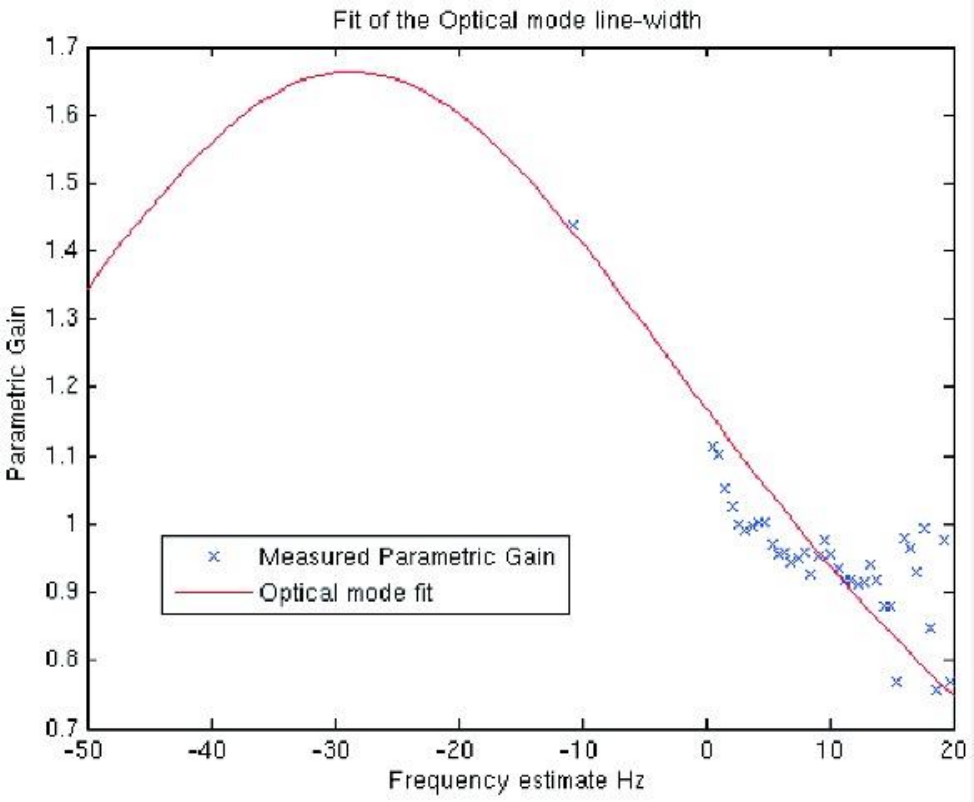}\vspace{-2mm}
\caption{(Color online) The parametric gain (PG) as a function of estimated mode frequency shift after a transient in the ring heater with a Lorentzian fit.  The gap in the data is due to the cavity unlocking due to parametric instability and then being relocked}\label{OpticalModes1}
\end{figure}

Figure~\ref{OpticalModes1} shows a Lorentzian fit that approximates the RoC and thus the HOOM frequency transient.  Although the measured PG did not pass peak in the sweep, the peak of the Lorentzian fit is $28$ Hz to the left of zero.  Based on this measurement the Y arm HOOM spacing, with $100$ kW contained power and a ring heater operating at $1.1$ W, can be estimated as $15032\substack{+20\\ -5}$ Hz$^{3)}$\footnote{3) Measurements done when ETMX ring heater was in a faulty condition prior to 9 July 2015} and the peak PG $\sim 1.7\pm 1$.  Using a similar method with different lock stretches the Y arm HOOM spacing has been estimated as $15400\substack{+70\\ -100}$ Hz with $100$ kW in the cavity and the ring heater operating at $0.8$ W and a predicted maximum PG of  $\sim8$.

\subsection{Overlap parameter}\label{Overlap}

\begin{figure}[H]
\centering
\includegraphics[scale=0.48]{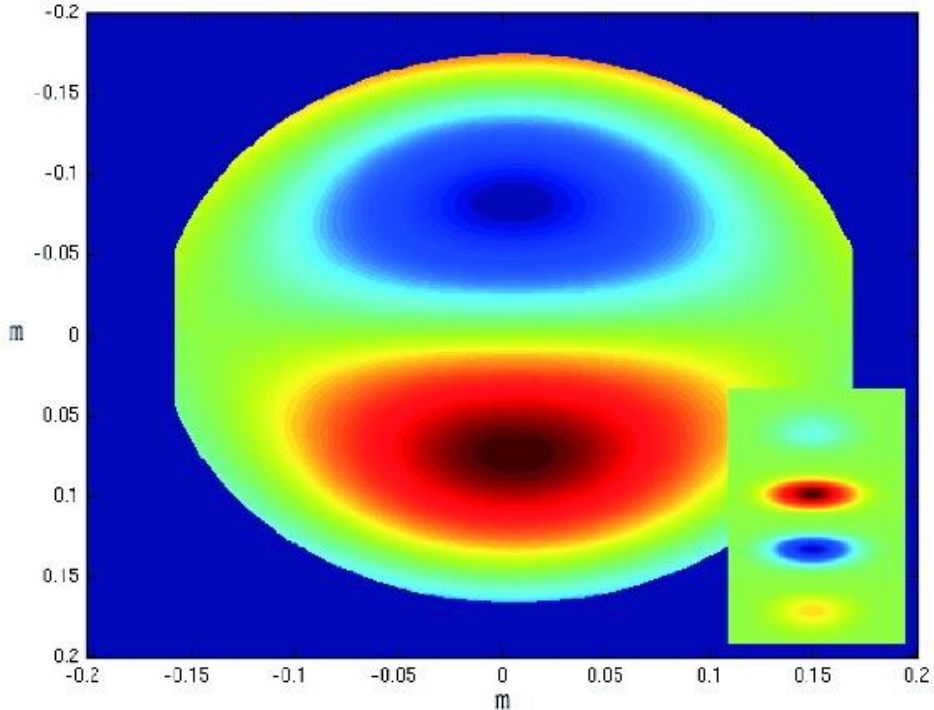}\vspace{-3mm}
\caption{(Color online) The acoustic mode surface deformation and beat note spatial profile (green inset) at the mirror used to estimate $b$. the normalized overlap function. } \label{OverlapProd}
\end{figure}

The overlap parameter $\Delta_{\rm s}$ in eq.~(\ref{parametricGain}) is defined as \cite{Braginsky}:
\begin{equation}
\Delta_{\rm s}= \frac{V \big(\int f_0({\bm r})f_1({\bm r})u_z\partial r \big)^2}{\int \abs{f_0}^2\d{\bm r}\int\abs{f_1}^2\d{\bm r}\int\abs{{\bm u}}^2}, \label{eq:overlap}
\end{equation}
where $f_0$ and $f_1$ are the optical field distributions over the mirrors surface shown as their product in Figure~\ref{OverlapProd} inset, $u_z$ is the surface deformation tangential to the surface shown in Figure~\ref{OverlapProd} ie the tangential component of ${\bm u}$, $\int \d{\bm r}$ are integrals over the surface and $\int \d{\bm V}$ integrals over the mirror volume

\begin{figure}[H]
\centering
\includegraphics[scale=0.55]{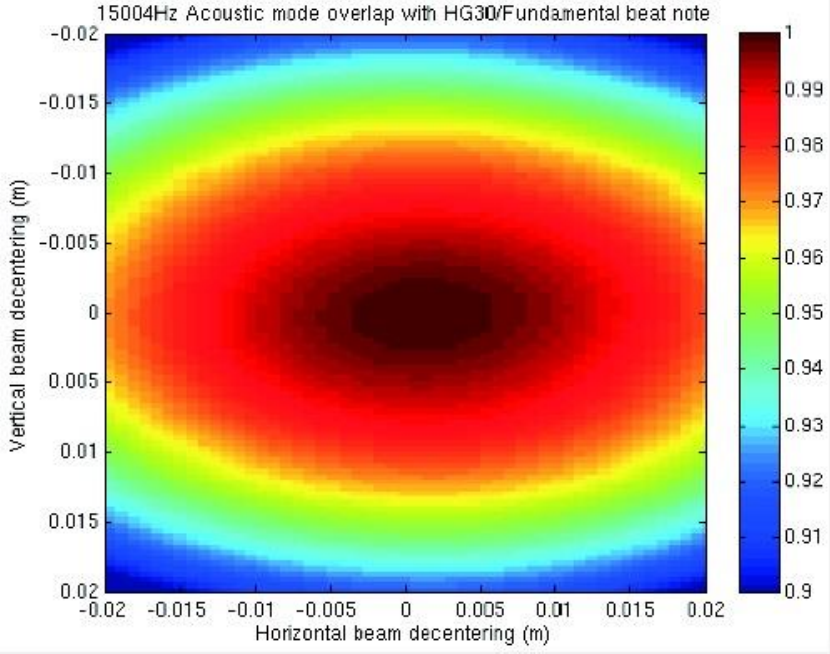}\vspace{-3mm}
\caption{(Color online) The normalised convolution of the optical beatnote and the mirror tangential surface deformation. }  \label{OverlapFunc}
\end{figure}

The expected overlap parameter can be calculated from simulated mode shapes assuming nominal alignment to be 0.11.  To determine if the overlap was likely to have experimental consequences a simulation of the relative overlap parameter was made for the 15538 Hz mode and an ideal $TEM_{03}$ and $TEM_{00}$ beat note by computing the numerator in eq.~(\ref{eq:overlap}) over the expected range of beam positions on the optics, as depicted in Figure~\ref{OverlapFunc}.

The normalized overlap as a function of position shows that during normal interferometer operation with less than $2$ mm variation in spot position and the estimated lock to lock variation of $\pm 10$ mm based on an estimate from the angle to length decoupling gain variations, the maximum variation in the overlap parameter is in the order of 2\% which is insignificant.

\subsection{Parametric gain}\label{Parametric_Gain}
The $Q$ of the acoustic modes was estimated from material \linebreak
\vspace{-2mm}

\begin{figure}[H]
\centering
\includegraphics[scale=1.1]{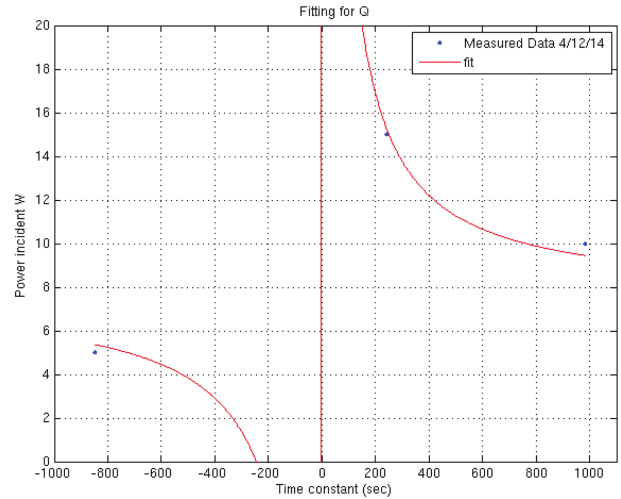}
\caption{(Color online) The fit used to estimate the $Q$ of the acoustic mode in the 15004 Hz mode with the measured data as the inset.  The estimated $Q$ is $\times 10^6$. }
\label{fig:tau2R}
\end{figure}

\noindent properties to be $\sim10$--$100\times10^6$ \cite{StriginNum}.  We have a  lower bound of $\sim 7 \times 10^6$ from linewidth measurements.  At the time of the linewidth experiments the electrostatic actuators were not set up to drive at the desired PI frequencies.   Further work on using electrostatic drives to damping PI would make it possible to excite and directly measure the ring down of acoustic modes.  Another method of obtaining the acoustic $Q$ was described by Miller et al. \cite{Electrostatic}, where several measurements of the parametric gain at different contained powers can be used to derive the $Q$ of the mechanical mode. By definition $R-1 = 2Q_{\rm m}/(\omega_{\rm m} \tau)$, and combined with eq.~(\ref{parametricGain}) the proportionality between $R$ and $P$ is
\begin{equation}
A\times P =  2Q_{\rm m}/(\omega_{\rm m} \tau +1). \label{eq:tau2R}
\end{equation}
Using two measurements of $\tau$ and $P$ we can solve for $Q_{\rm m}$ and the assumed constant parameter $A=\frac{McL\omega_{\rm m}^2(1+ \Delta_\omega/\delta_{\rm o})}{Q_{\rm m}Q_{\rm o}\Delta_{\rm s}}$.  Figure~\ref{fig:tau2R} shows a fit for mode 15004 Hz to obtain the $Q$ factor using eq.~(\ref{eq:tau2R}).

This method has been used to estimate the $Q$ of several modes. The $Q$-factors of those two modes responsible for instability, the 15538 Hz and the 15004 Hz are $(12 \pm 4) \times 10^6$ and $(6.5 \pm3) \times 10^6$, respectively.

We can now derive the expected PG for the simplified single cavity approximation using eq.~(\ref{parametricGain}).   Tthe parametric gain for the 15538 Hz mode is $7\pm3$ and for the 15004 Hz mode is $3\pm2$, for the case where the frequency overlap is maximum.  The Lorentzian fit in Figure~\ref{OpticalModes1} would predict a maximum PG of $4\pm1$,  so within tolerances we have some agreement.  However it must be noted that this is a heavily simplified single cavity model.  The model will need to be fine tuned in the likely situations where the signal recycling and power recycling cavities come into play, with significantly larger PGs, as shown in ref. \cite{EvensPIgeneral}.

\subsection{Summary}\label{sec_fin}
We have reviewed the physics of parametric instability and presented observations that largely confirm predictions made in 2005. Studies on the aLIGO interferometer at Livingston have demonstrated that parametric instability behaves as expected, but with large experimental uncertainties and a simplified single cavity model.  It showed that we have a good model and method for determining the acoustic mode frequencies in the aLIGO test masses.  Efforts are needed to improve the measurement tolerances on the $Q$ of the mechanical modes and the HOOM spacings, so that firm predictions of parametric gains at high power can be made.  It would appear that spatial overlap is of minimal concern when considering avoidance of parametric instability.  Currently experimental results would suggest that there is a space between the 15.0 kHz modes and the 15.54 kHz modes that appears to have relatively low interacting acoustic modes and a space from 15.3 kHz to 15.5 kHz with no acoustic modes.  This would suggest a wide possible tuning range and transient tolerance such that thermal tuning with ring heaters can be used for the avoidance of parametric instabilities at the modes around 15.0 kHz and 15.54 kHz.  Measuring the $Q$ factors of modes with relatively small parametric interactions will be essential to determining how safe this region is, and enabling an estimate of the maximal optical power thermal tuning alone will accommodate.

\section{Optomechnical devices for improving gravitational wave detector sensitivity}
\emph{Advanced GW detectors will be reaching the quantum noise limit in most frequencies of the detection band. We review the efforts in improving quantum noise limited sensitivity of  GW detectors using optomechanical devices. A novel mechanical resonator design combined with quantum noise-free optical dilution was proposed to mitigate the thermal noise in optomechanical devices. Theoretical sensitivity improvements were given using optomechancial narrow band filters for frequency dependent squeezing and white light cavities assuming that the thermal noise is negligible with optical dilution.}

\subsection{Introduction}\label{sec:intro}
GWs detectors must be able to measure almost infinitesimal vibrations so as to observe events such as the merger of pairs of black holes in the distant universe. The sensitivity of the first generation GW detectors such as LIGO reached the quantum-shot-noise limit in the high-frequency part of the spectrum. In the second-generation detectors now in commissioning, quantum radiation-pressure noise is expected to dominate at low frequency detection band, while shot noise will dominate at high frequencies. At the region around 100 Hz the sensitivity is limited by classical test mass thermal noise. But as better optical coatings and test masses become available, future detectors should be limited mostly by quantum noise.

A succession of techniques has been proposed over the last 30 years for breaking the quantum limit barrier. Two of the techniques are particularly promising: the optical squeezing \cite{Caves}, and the signal recycling \cite{Meers,Yanbei}. Both of these techniques have been proven, but both have significant bandwidth limitations.  The quantum uncertainty in phase space is squeezed such that one variable (i.e. that contributing to the shot noise) has reduced uncertainty while the other (i.e. contributing to the quantum radiation-pressure noise) has increased uncertainty, or verse versa.   Recently  optical squeezing \cite{LigoSQ, GeoSQ} has been spectacularly demonstrated. In 2001 Kimble et al. \cite{Kimble} showed that squeezing can be substantially enhanced with special optical filters that are used to rotate the squeezing angle in a frequency dependent (FD) way so that both shot noise and radiation– pressure noise can be reduced simultaneously, so as to achieve broadband sensitivity improvement. Chelkowski et al. \cite{Chelkowski} demonstrated a FD squeezed vacuum using a short filter cavity in the MHz range. Most recently, Oelker, et al. \cite{Oelker} demonstrated the FD squeezing around 1.2 kHz using a 2-m-long, high finesse optical cavity. We will discuss this further below.

Signal recycling is a significant tool in optimising interferometers. Depended on the specification of the signal recycling mirror parameters, it can be used to resonantly enhance the signal or the detection bandwidth, and it can also be used in a detuned manner to change the interferometer dynamics and thereby achieve sensitivity below the free mass standard quantum limit.  Unfortunately the signal recycling enhancement of the sideband light which carries the GW signal only allows optimisation at a single frequency. If the recycling cavity losses are reduced, sensitivity is improved in a very narrow bandwidth. Thus signal recycling makes it difficult to retain broad bandwidth. Wicht et al. \cite{Wicht} have shown that this problem can be overcome through use of negative dispersion to create a white light signal recycling cavity.

The white light cavity uses the phase advance created by the negative dispersion medium to compensate the phase lag of the cavity round trip of the light. In this case, the cavity is almost resonance at all sideband frequencies. The white light cavity provides great benefit for signal recycling interferometer detectors. The traditional signal recycling increases the sensitivity at specific frequency by sacrificing the detection bandwidth.  The white light signal recycling cavity can have increased sensitivity at broadband frequencies.

\begin{figure}[H]
  \centering
  \includegraphics{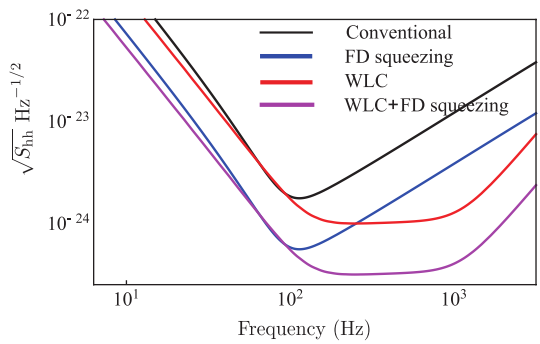} \vspace{-2mm}
  \caption{(Color online) The quantum noise limited sensitivity of various interferometer configurations, black line: a conventional interferometer; blue line: with frequency dependent (FD) squeezed vacuum injection at the dark port; red line: with white light cavity (WLC) signal recycling; pink line: white light cavity signal recycling combined with FD squeezing.}
  \label{fig:figure 1}
\end{figure}

If both of the above schemes could be implemented together, an interferometer could in principle access substantially improved sensitivity as shown in Figure \ref{fig:figure 1}. Frequency dependent squeezing can be implemented using extremely high performance conventional optical cavities. However white light cavity technology cannot be created using conventional materials. The use of optically pumped atomic system was first proposed to be used in optical cavities to create narrow band filters \cite{Mikhailov2006} as well as white light cavities \cite{Wicht}.

\noindent However, the active atomic system is generally lossy, which will limit the sensitivity improvement \cite{MaQN}.

In this section we discuss methods whereby ultra-low noise optomechanical cavities can be used to create both compact filter cavities and white light cavities. Sect. 6.2 discusses the narrow band filter for frequency dependent squeezing using optomechanical cavity; sect. 6.3 discusses the negative dispersion created by optomechanical cavity and theoretical sensitivity improvement by using it in white light signal recycling cavity; sect. 6.4 discusses a novel way to create an ultra-low thermal noise mechanical resonator for optomechanical devices; sect. 6.5 gives a summary.

\subsection{Optomechanical narrow band filters}

In order to achieve frequency dependent squeezing and hence broadband reduction of quantum noise in an interferometer detector, the frequency scale of the filter cavities needs to match that of quantum noise of the main interferometer. For the aLIGO detectors, the quantum noise is dominated by quantum radiation pressure noise at low frequencies and shot noise at high frequencies. The transition from radiation pressure dominance to the shot noise dominance happens around 100 Hz, which determines the required filter-cavity bandwidth.

Kimbles original proposal \cite{Kimble} used filter cavities of kilometre length. Recently, Evans et al. \cite{Evans} proposed a more compact (10 m) filter cavity with $10^5$ finesse to achieve the required bandwidth. With such a high finesse, even small optical losses can degrade the squeezing. Therefore, the optical loss becomes the key limiting factor in the filer-cavity performance. Isogai et al.\cite{Isogai} have experimentally demonstrated that the optical losses from current mirror technology are sufficiently small to build such a filter cavity that will be useful for the aLIGO detector \cite{aLIGO}. However, for future detectors of even greater sensitivity, with higher levels of squeezing, the traditional simple optical cavity may not meet the requirement. An alternative approach that can create a flexible and tunable ultra-compact filter cavity, is to use an optomechanical filter cavity.

Optomechanical devices make use of strong interactions between light (or electromagnetic waves in general) and mechanically resonant mirrors in optical cavities. The technology is progressing fast. Mechanical motion is strongly coupled to a resonant light field by radiation pressure. In such systems the phonons and photons become effectively mixed, and particular tuning conditions allow surprising optical responses analogous to that of atomic systems.

One of the phenomena that can be utilised using optomechanical cavities, called optomechanically induced transparency, enables the  creation of a cavity linewidth that can be as small as the linewidth of the mechanical resonator involved. It can be tuneable by tuning the optomechanical coupling strength with variable laser power. Such an optomechanical cavity can be used as an optical filter cavity for squeezing angle rotation as required for creation of frequency dependent squeezing as described below.

In principle, the optomechanical cavity can be made to be on the centimetre scale while still having a bandwidth comparable to a much longer high finesse cavity. Additionally, with an active element, the cavity optical properties can be dynamically tuned by changing the power of the control pumping field. This has the advantage of allowing optimization of the filter cavity for different operational modes of the detector, where the quantum noise has different frequency dependencies. For example it could allow aLIGO-type detectors to be switched between tuned and detuned resonant sideband extraction schemes, to optimise detectability of certain GW signals.

Optomechanical filter cavities based on optomechanically induced transparency have been studied and demonstrated by various research groups \cite{Weis, Teufel,Safavi}. Ma et al. \cite{MaFilter} proposed that the optomechanical filter cavity can be used for frequency dependent squeezing. Qin et al. \cite{QinFilter} showed that this effect could be used to achieve tunable narrow linewidth cavities.

\begin{figure}[H]
  \centering
  \includegraphics{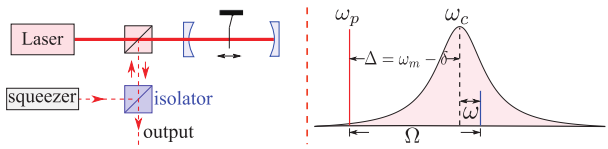}
  \caption{(Color online) Experimental concept of an optomechanical filter cavity (left) and the frequency diagram with all the frequencies involved (right) with the mechanical resonator frequency $\omega_{\rm m}$, the cavity resonant frequency  $\omega_{\rm c}$  and the control beam at frequency $\omega_{\rm p}$ which is red detuned from the cavity frequency by $\omega_{\rm m}-\delta$.}
  \label{fig:figure 2}
\end{figure}

Figure \ref{fig:figure 2} shows the concept of the filter cavity. The mechanical resonator with resonant frequency $\omega_{\rm m}$ is placed inside an optical cavity of resonant frequency $\omega_{\rm c}$. The value of $\omega_{\rm m}$ is arbitrary because the pumping frequency can always be detuned such that the final filter frequency has the chosen value. The control laser beam is red detuned from the cavity resonance by $\omega_{\rm m}-\delta$. Using standard vacuum squeezing technology, the squeezed vacuum at frequencies close to the cavity resonance is injected into the cavity. The optomechanical interaction between the strong control beam, the squeezed vacuum and the mechanical resonator causes the squeezed vacuum reflected from the cavity to experience an effective narrow band filtering. The squeezing angle rotates as a function of the frequency detuning from the effective cavity resonance frequency.

\begin{figure}[H]
  \centering
  \includegraphics{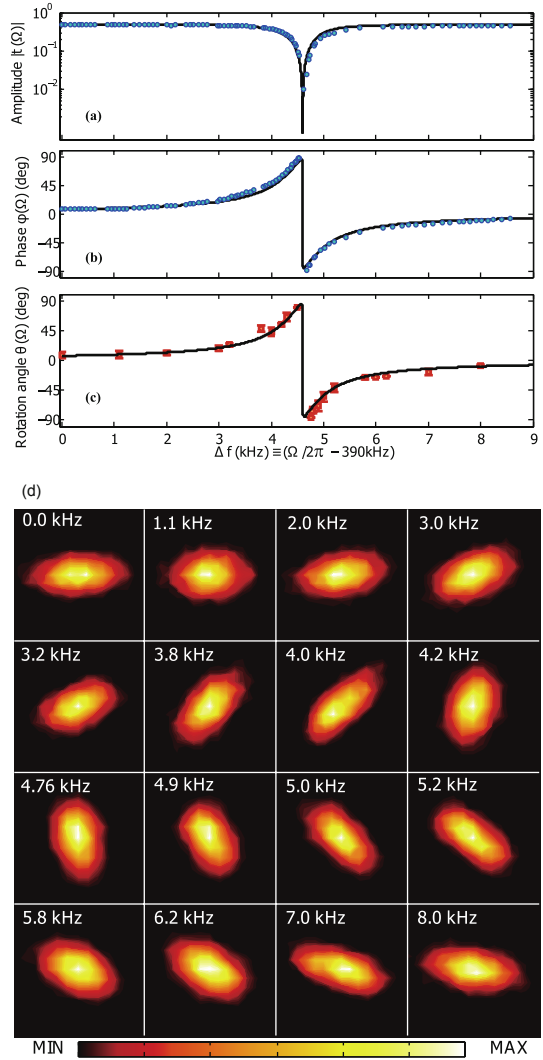}
  \caption{(Color online) Rotation of a classical noise ellipse using a membrane-in-the-middle cavity. (a) Transmitted amplitude; (b) phase; (c) rotation angle; (d) contour plotted phaser diagrams. Horizontal axis is amplitude quadrature, vertical is phase, obtained with frequency offset $\delta=4.6$ kHz.}
  \label{fig:figure 3}
\end{figure}

Figures \ref{fig:figure 3}(a) and (b) show the experimental results of filter cavity with tuneable narrow linewidths down to 2.6 Hz
\cite{QinFilter}. This is used to~ rotate a~ ``squashed'' light,~ which~ is a

\noindent classical imitation of a quantum squeezed light, in a frequency dependent way as shown in Figure \ref{fig:figure 3}(d). The results are still far from the quantum regime due to the high thermal noise of the mechanical resonator. However, if this technology is combined with optical dilution and optical cooling, as discussed in the following section, such a system could become a practical ultra-narrow band optical filter suitable for broadband squeezed light enhancement of GW detectors.

\subsection{White light cavity}
The idea of the white light cavity was first proposed by Wicht et al. \cite{Wicht} and considered by Salit and Shahriar \cite{Salit} in 2010 as a means of improving the sensitivity of the interferometric GW detectors by enabling a broad band of frequencies to be simultaneously resonant and therefore obtaining an enhanced sensitivity. This is a revolutionary idea.  A white light cavity breaks the nexus between bandwidth and resonant amplification that is fundamental to classical systems. The original idea was to use a pumped three-level atomic gas system to generate a negative dispersion medium to be inserted into the cavity. Because lower frequencies travel faster than high frequencies both can be simultaneously resonant. Thus with such a cavity a detector can benefit from resonant signal build-up across a broad band of frequencies. Wicht's \cite{Wicht} and Salit's \cite{Salit} analysis showed great enhancement of the signal response but did not consider quantum noise nor the signal to noise ratio.  Pati et al. \cite{Pati} experimentally demonstrated an atomic system white light cavity in the classical regime.

Until recently, the white light cavity was thought to be unusable for GW detectors because the quantum noise due to the intrinsic loss of the atomic system, and the stability criterion put such a stringent constraint on the behaviour of the interferometer configuration that it will destroy any benefit of the signal response enhancement as shown by Ma et al. \cite{MaQN}. However, Miao et al. \cite{MiaUnstable} realised that the physics of atomic interactions can be replaced by optomechanical interactions, for which losses can be minimal.

Miao et al. showed that a set up similar to that shown in Figure 30, but with a blue detuned control laser frequency set to
$\omega_{\rm c} +\omega_{\rm m}$ with the mechanical oscillator frequency $\omega_{\rm m}$ much larger than the cavity bandwidth. The cavity resonance at $\omega_{\rm c}$ is in favour of the down-conversion process, which amplifies the sidebands around $\omega_{\rm c}$ due to the mechanical motion. It can be viewed as a phase-insensitive parametric amplifier for sidebands. The pumping of the optomechanical filter in the unstable (blue-detuned) regime gives rise to the optical negative damping which is much larger than the resonator internal damping. It is critical to use feedback control to stabilise the system. In a practical implementation, the feedback signal contains both the GW signal together with noise, such that the feedback does not influence the signal to noise ratio. The analysis by Miao et al. \cite{MiaUnstable} showed that the optomecahnical filter exhibits negative dispersion.  When an appropriately tuned negative dispersion optomechanical filter is placed inside the signal recycling cavity, the cavity round-trip phase lag is compensated, and therefore broadband resonance enhancement in sensitivity is expected to be achieved as shown in the red curve of Figure \ref{fig:figure 1}.

Most recently, Qin et al. \cite{QinNeg} analysed a self-stabilised blue detuned optomechanical filter. It was shown that this system can be used in white light signal recycling cavity to enhance the sensitivity-bandwidth product with a detuned signal recycling cavity for specific GW signal frequencies \cite{QinPre}.

\begin{figure}[H]
  \centering
  \includegraphics{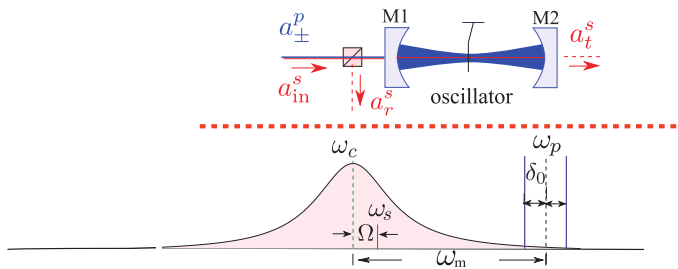}\vspace*{-2mm}
  \caption{(Color online) Schematics for a negative dispersion optomechanical cavity. The resonant frequency of the cavity is $\omega_{\rm c}$. The position of the membrane is chosen to introduce a linear optomechanical coupling. The radiation pressure forces from the beating between the signal light at the frequency $\omega_{\rm p}-\omega_{\rm s}=\omega_{\rm m}-\Omega \pm \delta_0$ and the control beams at the frequencies $\omega_{\rm p}=\omega_{\rm m}\pm\delta_0$ drive the mechanical oscillator which in turn creates sidebands that destructively interfere with the signal light, thereby creating linear negative dispersion.}
  \label{fig:figure 4}
\end{figure}

In 2014, Qin et al. \cite{QinNeg} experimentally demonstrated an optomechanical system where a tuneable linear negative dispersion was created with a blue detuned doublet of control beams as shown in Figure \ref{fig:figure 4}. In this experiment, the air damping was used to increase the mechanical resonator linewidth to maintain the system stable but with higher thermal noise. In future application, the optical damping could be used with relatively low noise injection contributed by the low-temperature optical bath. The mechanical resonator consisted of a silicon nitride membrane of resonance frequency $\omega_{\rm m}$ placed in the middle of a cavity. Two control beams $2\delta_0$ apart in frequency are blue detuned from the cavity resonance frequency $\omega_{\rm c}$ by $\omega_{\rm p}=\omega_{\rm m}\pm\delta_0$. The signal beam at frequency $\omega_{\rm s}=\omega_{\rm c}+\Omega$ beats with control beams to create radiation pressure forces driving the mechanical resonator. Its motion scatters the control beam into sidebands at the signal frequency and constructively interferes with the signal beam. The signal beam reflected from the cavity experiences negative dispersion due to this optomechanical interaction. The system could have very low optical losses dominated by a few ppm optical coating loss \cite{OptLoss}.

\begin{figure}[H]
  \centering
  \includegraphics[scale=0.38]{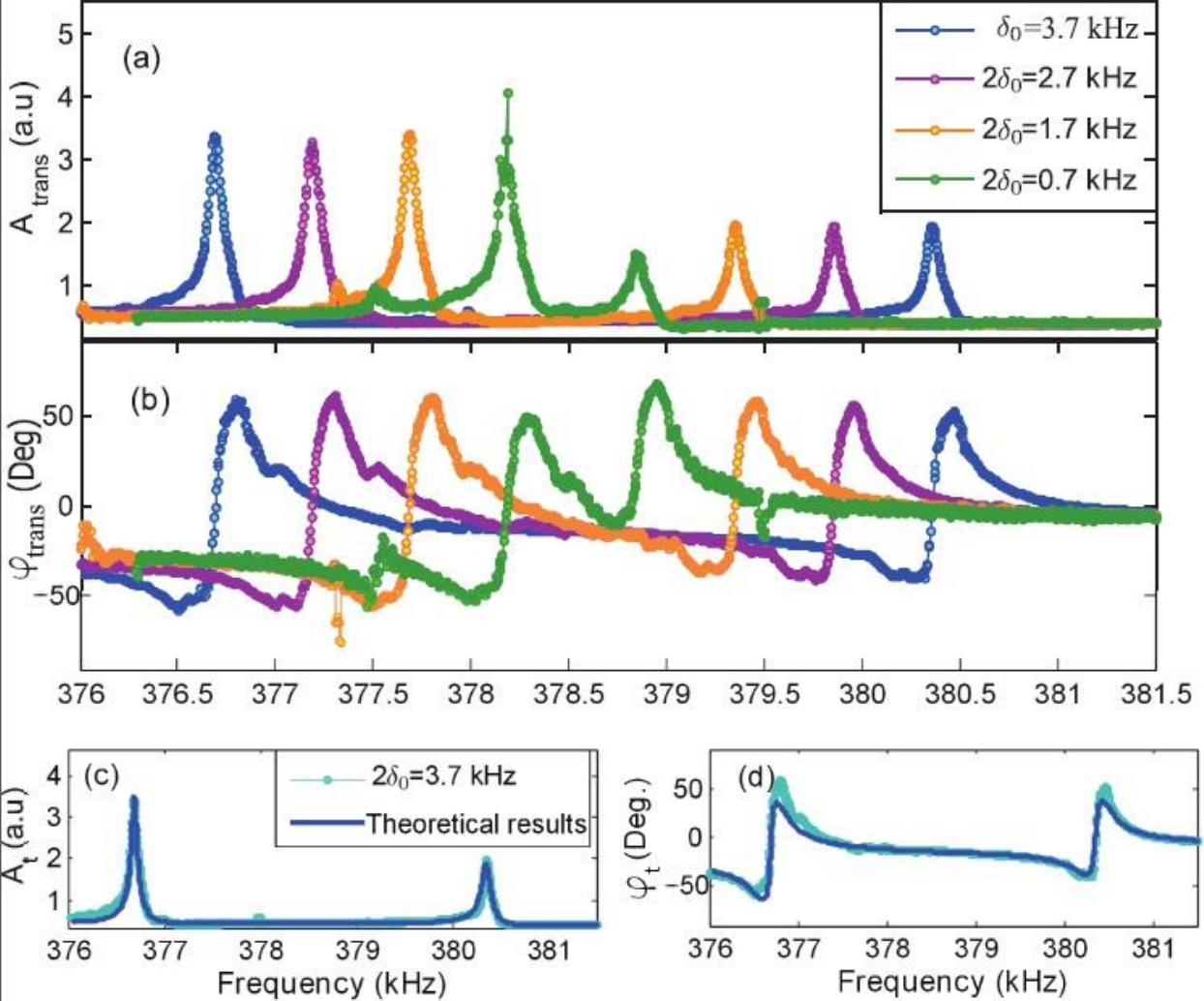}
  \caption{(Color online) Results from Qin et al. experiments regarding negative dispersion in a membrane-in-the-middle cavity \cite{Wicht}. (a) Transmitted amplitude, (b) transmitted phase, featuring a negative gradient versus frequency, (c) and (d) theoretical vs. experimental results for a frequency offset of 3.7 kHz.}
  \label{fig:figure 5}
\end{figure}

Figure \ref{fig:figure 5} shows experimental results for negative dispersion, achieved using various frequency gaps between the two control beams. It is clear that the phase decreases with increasing frequency in the central region (Figures \ref{fig:figure 5}(b) and (d)). The slope becomes small and has a broad linear region as the control beam frequency gap increases. There are two peaks of gain greater than unity in amplitude (Figures \ref{fig:figure 5}(a) and (c)) corresponding to two control beam frequencies. It is necessary to suppress these gains to prevent instabilities. Because the gain peaks are outside the linear dispersion band, it is not hard to restrict the feedback control outside the linear dispersion band without introduce noises.

\begin{figure}[H]
  \centering
  \includegraphics{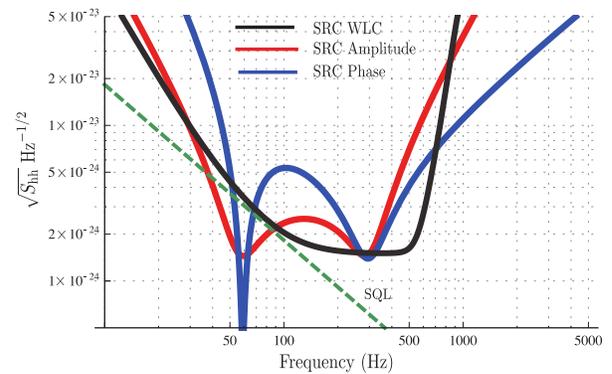}
  \caption{(Color online) The quantum noise limited sensitivity curves show the improvement of white light cavity: dashed line: free mass standard quantum limit (SQL); blue line: the sensitivity by measuring the amplitude quadrature of the detuned signal recycling interferometer output; red line: the sensitivity by measuring the phase quadrature of the detuned signal recycling interferometer output; black line: the sensitivity of the white light cavity signal recycling configuration.}
  \label{fig:figure 6}
\end{figure}

Figure \ref{fig:figure 6} shows the sensitivity improvement if optomechanically created negative dispersion is used to create a white light signal recycling cavity in aLIGO type detectors. The dashed line is the free mass standard quantum limit. The blue and red curves are detuned signal recycling interferometer sensitivity in phase and amplitude quadrature. There are two dips in the sensitivity curves. The dip at the high frequency corresponds to the cavity resonance. The dip at the low frequency corresponds to the optical spring shifted mechanical resonance. The black curve is the sensitivity with stable white light signal recycling cavity. Because of the broadband resonance of the white light cavity, the optical spring effect disappears and so does the low frequency dip, while the high frequency dip expands.

The sensitivity improvement discussed above considered only quantum noise sources. With non-zero environment temperature, the mechanical resonator will be thermally driven. The thermal motion of the resonator will then scatters the control light into the signal light as noise. In order for the thermal noise to be lower than the quantum shot noise, we require \cite{MiaUnstable}  $8k_{\rm B}T/Q_{\rm m}\leq\hbar\gamma$, where $T$ is the environment temperature; $Q_{\rm m}$ is the mechanical resonator $Q$-factor; and $\gamma$ is the effective cavity bandwidth, which is equal to the optomechanical anti-damping rate.  As an order of magnitude estimation, we have, $T/Q_{\rm m}\leq6\times10^{-10} \gamma/(2\pi\times100$ Hz). This is a strict requirement on environment temperature and the quality factor. In the following section we present a novel mechanical resonator design capable of very high optical dilution factors that could enable the benefits of opto-mechanics to be realised in future GW detectors.

\subsection{Ultra-low thermal noise mechanical resonators through optical dilution}

As shown above, optomechanical devices can potentially create narrow band optical filters and linear negative dispersion for creating the white light signal recycling cavities. The combined techniques could lead to substantial  sensitivity enhancement as shown in Figure \ref{fig:figure 1}. However none of this will be possible unless the thermal noise of the mechanical resonator is suppressed below the quantum noise.

One potential approach to ultra-low thermal noise optomechanics is to use quantum-noise-free dilution. The dilution of acoustic losses by non-dissipative springs is well known. For example gravitational restoring forces allow the quality factor of a pendulum to exceed that of the flexure from which it is suspended. The position dependence of radiation pressure created by optical standing waves creates optical traps or optical springs. If the optical spring significantly exceeds the stiffness of the mechanical spring the effect of thermal fluctuations is diluted. The dilution factor is the ratio of the elastic energy stored in the optical field to the elastic energy stored in the mechanical spring.

The feasibility of strong optical dilution was demonstrated by Corbitt et al. \cite{Corbitt} in 2007. A simple optical spring was used to raise the frequency of a 1gram pendulum from 12 Hz to 1 kHz corresponding to a dilution factor $\sim10^4$. Dilution increases as the square of the optical spring induced mechanical frequency as follows: $Q_{\rm dil}=Q_{\rm int}\frac{\omega^2_{\rm dil}}{\omega^2_{\rm int}}$. Analysis shows that much larger dilution factors  are in principle attainable, sufficient to increase the $Q$-factor above the limits discussed in the previous section. However there is an additional problem that must also be addressed. The problem with simple optical springs is the contribution of quantum radiation pressure noise. Strong dilution requires a strong optical field acting on the mechanical resonator. By beating with vacuum fluctuations this creates strong radiation pressure noise which drives the resonator, thereby injecting extra noises and setting limits on the maximum dilution \cite{MaFilter}.
There are several method proposed to solve this problem by achieving a quantum-noise-free optical diluition. Chang et al. \cite{Chang} considered the optical dilution limits for a membrane micro-mirror trapped by a cavity standing wave, which belongs to the so-called trapping through quadratic coupling. Key parameters are the mirror mass to suspension mass ratio, and mirror deformation from the intensity distribution of light on the mirror. Members of same team, led by Kimble achieved an important proof of principle in 2012 \cite{Ni}, demonstrating an optical trapping that achieved 145 kHz, and a 50-fold increase in quality factor, consistent with Chang''s prediction \cite{Chang}. The mirror to flexure mass ratio limited their $Q$-factor, while torsional compliance and internal acoustic modes allowed optically induced angular instabilities. Besides, Korth et al. \cite{Korth2013} proposed to use double optical spring scheme which evades the radiation pressure noise through sensing and feedback control.

Similar to ref. \cite{Ni}, Ma et al. \cite{MaFilter} analysed a dilution scheme where the resonator mirror sits in the middle of an optical cavity that is similar to the dispersive cavity cooling scheme \cite{Harris}. The detailed analysis showed that a high reflectivity end mirror enables total destructive quantum interference to cancel the radiation pressure noise \cite{MaFilter}.  Instabilities from negative damping are also cancelled within an attainable parameter range.
Subsequently the UWA team has shown that a second topologically equivalent configuration using a double sided high reflectivity mirror has the same performance \cite{Page}.

Noise free dilution not only increases the resonator frequency but also reduces the interaction strength with the thermal bath, thereby increasing the signal to noise ratio for detecting weak forces in a defined bandwidth.
To achieve the high degree of optical dilution requires resonators with the lowest possible intrinsic mechanical frequency. One possibility is to use completely levitated mechanical resonators \cite{ANU}. To reduce the number of degrees of freedom, a simpler solution can be a pendulum suspended by an extremely low rigidity membrane or wires.

\begin{figure}[H]
  \centering
  \includegraphics[scale=0.45]{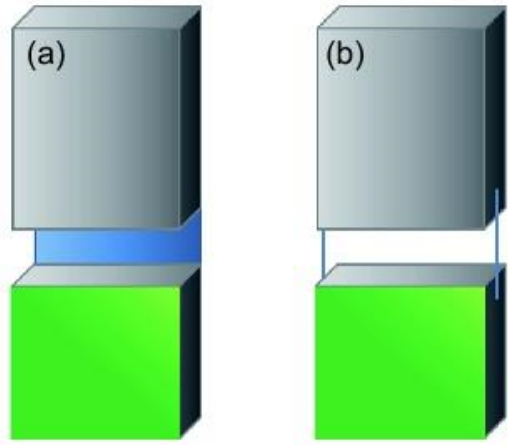}\vspace*{-2mm}
  \caption{(Color online) Cat-flap mirror designs. The bottom part is the mirror with high reflectivity coating on both sides, which is suspended from top base by, (a) a membrane; (b) two nanotubes.}
  \label{fig:figure 7}
\end{figure}

We have devised a membrane suspended pendulum, also referred to as a "cat-flap" resonator, which has minimal coupling to the thermal environment. The concept is shown in Figure \ref{fig:figure 7}, with the membrane suspension being made of silicon nitride, graphene or carbon nanotubes.

The suspended mass has a high reflectivity dielectric coating on both faces. The minimisation of the ratio between the mass of the tether and pendulum bob has been shown by Chang to be beneficial to dilution \cite{Chang}. Modelling of the suspension mechanism of the resonator is ongoing, with one of the most important considerations being the frequency of membrane internal modes that significantly reduce the quality factor obtained by dilution as seen by Ni et al. \cite{Ni}.

\begin{figure}[H]
  \centering
  \includegraphics{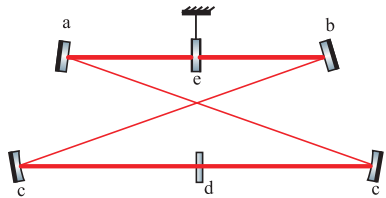}
  \caption{(Color online) Bowtie cavity that is equivalent to a membrane-in-the-middle coupled cavity. a Input mirror, $T\sim100$ ppm; b flat mirror, $T\sim10$ ppm; c curved mirrors, $T\sim10$ ppm; d cat-flap resonator; e sloshing mirror $T\sim100$ ppm. }
  \label{fig:figure 8}
\end{figure}

A double sided high reflectivity mirror can be used in the quantum radiation pressure noise cancelling configuration analysed by Ma et al. \cite{MaFilter} as long as it is combined with a cavity coupling mirror. We propose that the cat-flap mirror be used in a ring cavity configuration shown in Figure \ref{fig:figure 8}. This cavity exactly maps to the coupled membrane-in-the-middle cavity. The cat-flap mirror is topologically equivalent to a pair of end mirrors, while the coupling mirror (also called a sloshing mirror) replaces the function of the partially transmissive membrane.

To achieve sufficient thermal noise suppression, all the loss mechanisms for optical spring resonators must be addressed.  These include acceleration losses, losses through the pendulum suspension, and thermoelastic losses. Recoil loss can be mitigated by aligning the optical spring beam at the resonator's centre of percussion. Losses from optical spring coupling to internal modes is minimised by using small resonator with relatively large laser spot. Another source of dissipation is that the oscillation of the pendulum causes slight movement of the laser spot relative to the surface, causing temperature gradients and internal motion that limits the effects of the dilution. This effect is proportional to the upper optical spring frequency, limiting the amount of dilution that can be achieved.

We are currently considering various suspension options that include graphene, silicon nitride, carbon nanotubes and silicon nanowires. The micromirror size must be optimised to keep internal acoustic modes as high as possible, while optical diffraction loss minimisation prevents the mirror size from becoming too small.

\subsection{Conclusions}

In summary, optomechanical devices discussed here will provide options for future improvements of advanced detectors\linebreak
\vspace*{-3mm}

\noindent \!\cite{LIGOby}, and for the 3rd generation GW detectors such as the ET \cite{4.1ET}. The major obstacle for these active filters to work in the quantum regime is the thermal noise of the mechanical resonator. The proposed mechanical resonator design combined with noise-free dilution will provide the potential solution.

\section{Techniques for obtaining enhanced sensitivity in quantum-limited advanced gravitational wave detectors\ \ \ \ \ \ \ \ \ \ }

\emph{This part gives an overview of existing ideas on improving the sensitivity of advanced gravitational-wave detectors, emphasizing particularly the quantum noise. These ideas generally involve (i) modifying the input/output optics of the standard Advanced-LIGO-type power- and signal-recycled (dual-recycled) Michelson laser interferometer: e.g., frequency-dependent squeezing, frequency-dependent readout, sloshing-cavity speedmeter, and long signal recycling cavity, or (ii) introducing additional optical degrees of freedom: e.g., dual carrier scheme, or (iii) various combinations among them. We will summarize their key features, some further technical details behind this summary are contained in ref. [Class Quantum Grav, 2014, 31: 165010].}

\subsection{Introduction}\label{sec:introduction}

\begin{figure}[H]
  \centering
  \includegraphics{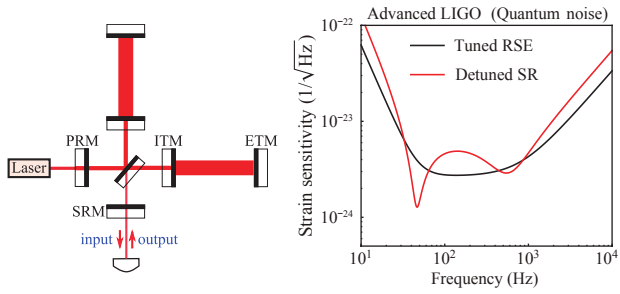}
\caption{(Color online) The dual-recycled Michelson interferometer (left) and quantum noise curves given two different SR detunings (right). Interferometer measures the differential motion of input test mass (ITM) and end test mass (ETM) induced by GWs. The vacuum fluctuation enters the dark port as input; the photodetector measures the output field via homodyne detection.}
\label{fig:config_aLIGO}
\end{figure}

One of the limiting noises of advanced GW detectors, including aLIGO\,\cite{4.1aLIGO}, Advanced VIRGO\,\cite{VIRGO} and KAGRA\,\cite{7Somiya} is the quantum noise that arises from quantum fluctuation of the optical field. The fluctuation in optical amplitude, when beating with the strong carrier field, produces a random radiation pressure on the mirror-endowed test mass, which gives rise to the so-called quantum radiation-pressure noise. It is directly proportional to the optical power inside the arm cavity. Due to the mechanical response of the test mass, the radiation-pressure noise is mainly dominated at low-frequency end of the noise spectrum. While at high frequencies, the so-called shot noise dominates, which arises from the quantum fluctuation of optical phase. In contrast to the radiation-pressure noise, it is inversely proportional to the optical power and has a nearly flat spectrum up to the detector bandwidth where the optical response starts to decrease. The transition frequency from the radiation-pressure noise dominant regime to the shot-noise dominant regime depends on the optical power and size of the test mass. Take aLIGO for instance. Such a frequency happens to be around 100 Hz and can be shifted around by a factor of few via tuning the signal-recycling cavity. The nominal operational mode of aLIGO is the so-called resonant-sideband-extraction (RSE) mode for enhancing the detector bandwidth, and the corresponding transition frequency is decreased to be around 70 Hz. The standard power-recycled (PR) and signal-recycled (SR) Michelson configuration of aLIGO and its quantum noise curve are illustrated in Figure \ref{fig:config_aLIGO} where\linebreak
\vspace*{-3mm}

\noindent both RSE and detuned signal recycling (SR) scenarios are shown.

\subsection{Configurations}\label{sec:configurations}

There have been extensive efforts in the GW community trying to reduce the quantum noise, in order to extend the cosmic reach of the GW detectors. Here we are going to discuss some existing ideas that have been studied in depth. One can refer to recent review articles for more detailed discussions\,\cite{Danilishin2012, haixing}. They generally involve (i) modification of input/output optics or (ii) introducing additional optical degrees of freedom. Here input and output are all referred to the differential (dark) port of the laser interferometer, as it is where the quantum fluctuation of the vacuum field enters (input) and also where the signal is extracted (output). For modification of input/output optics, there are (1) frequency-dependent squeezing [subsection \ref{subsec:FDS}],
(2) frequency-dependent readout [subsection \ref{subsec:FDR}], (3) sloshing-cavity speed meter [subsection \ref{subsec:SM}], (4) long signal recycling cavity [subsection \ref{subsec:LSR}]. For introducing additional degrees of freedom, there is the dual-carrier scheme [subsection \ref{subsec:DC}]. We will discuss each of them in this section.

\subsubsection{Frequency-dependent squeezing}\label{subsec:FDS}

The first approach for reducing quantum noise is to inject squeezed light, which has lower quantum fluctuation in either amplitude or phase, into the differential port. However due to the frequency dependence of the quantum noise, reducing noise over a broad frequency band requires the squeezing angle to also have proper frequency dependence---amplitude squeezed at low frequencies and phase squeezed at high frequencies. This is realized by using one optical cavity or several in cascade to filter the frequency-independent squeezed light generated from nonlinear crystals\,\cite{Kimble}. The number of filter cavities and their parameters for creating the matched frequency dependence is discussed in details in the appendix of ref.\,\cite{Purdue2}. As a rule of thumb, the number of filter cavities is roughly equal to the number of pole pairs (in Fourier/Laplacian domain) of the interferometer response function. For example, the dual-recycled Michelson interferometer contains one pair of poles from the test mass response and one pair from the arm cavity mode (using single-mode approximation that is valid when considering frequencies lower than one free spectral range of the arm cavity), and therefore two filter cavities, in principle, is required, as discussed in great details in ref.\,\cite{HaEA2003}.

\begin{figure}[H]
  \centering
  \includegraphics{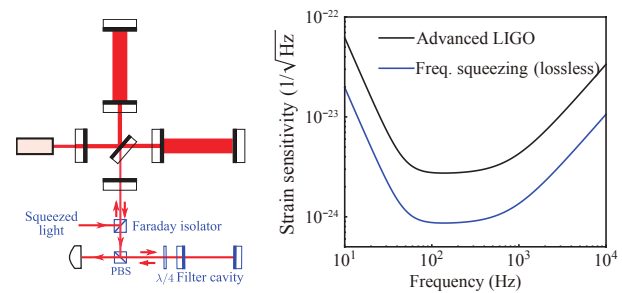}
\caption{(Color online) Schematics for realizing frequency-dependent squeezing with a filter cavity together with a squeezed light source.}
\label{fig:config_Sh_FDS}
\end{figure}

One interesting aspect noticed by Khalili\,\cite{Khalili2010} is that in the RSE mode, the cavity pole is at a much higher frequency than the transition frequency from the radiation-pressure noise dominant regime to the shot-noise dominant regime---one filter cavity can approximately provide the desired frequency dependence, which motivates Evans et al. \,\cite{Evans} to propose using a single 16 m linear optical cavity to create frequency-dependent squeezing for aLIGO, which is shown in the schematics on the left of Figure \,\ref{fig:config_Sh_FDS}.

In the ideal case, frequency-dependent squeezing allows for broadband reduction of the quantum noise by the squeezing factor as shown by the noise curve in Figure\,\ref{fig:config_Sh_FDS}.
One key constraint on its performance is the optical loss, from various sources, that degrades the squeezing. Recently, Kwee et al.\,\cite{Kwee2014a} made a systematic study of the optical loss in the frequency-dependent squeezing scheme. It was shown that the optical loss inside the filter cavity plays the dominant role in degrading the low-frequency part of the squeezing spectrum. This can be understood from the fact that high-frequency components of the squeezing field barely enter the filter cavity; while the loss effect on the low frequency part is directly amplified by a factor of the cavity finesse. As mentioned earlier, in order to match the frequency dependence of the quantum noise, the filter cavity bandwidth needs to be comparable to the quantum noise transition frequency, which is 70 Hz for aLIGO in the RSE mode. Given this fixed requirement, the finesse is inversely proportional to the cavity length---the shorter the cavity the higher the finesse, and the more important is the optical loss of the cavity. As pointed out in ref.\,\cite{Khalili2010}, optical loss per unit cavity length provides a good figure of merit for determining the filter cavity performance at low frequencies. Usually, the optical loss has a rather mild length dependence (cf. Figure 4 in \linebreak
\vspace*{-3mm}

\noindent ref.\,\cite{Evans}), which means a longer filter cavity generally can make the effect of optical loss smaller.

\subsubsection{Frequency-dependent readout}\label{subsec:FDR}

\begin{figure}[H]
  \centering
  \includegraphics{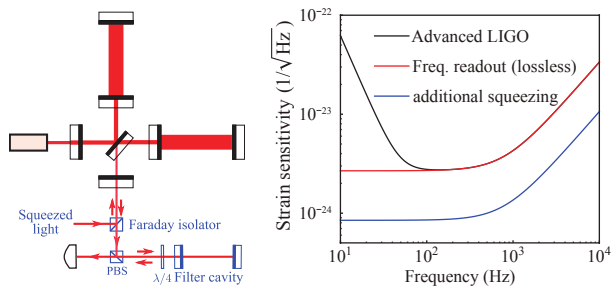}
         \caption{(Color online) The frequency-dependent readout scheme (left) which uses a filter cavity to rotate the readout quadrature in a frequency-dependent way to cancel the low-frequency radiation-pressure noise. Introducing an additional constant squeezing allows for a further reduction of the shot noise  (right).}
   \label{fig:config_Sh_FDR}
\end{figure}

The frequency-dependent readout is very similar to frequency-dependent squeezing---the former is filtering the output field while the latter filters the input field, as shown schematically in Figure\,\ref{fig:config_Sh_FDR}. In the ideal lossless case, it allows for a total cancellation of the low-frequency radiation-pressure noise and gives rise to shot-noise-only sensitivity. By merely increasing the power or using additional squeezing, the sensitivity can increase proportionally (see the noise curve in Figure\,\ref{fig:config_Sh_FDR}). However, it is quite susceptible to optical loss and also variation of the filter cavity parameters, because it works by cancelling the radiation-pressure noise at a price of sacrificing the signal response in the meantime. Only if the optical loss can be significantly reduced or the filter cavity length be increased, the frequency-dependent readout with constant squeezing in principle allows for a much better enhancement at low frequencies than the frequency-dependent squeezing.

\subsubsection{Sloshing-cavity speed meter}\label{subsec:SM}

\begin{figure}[H]
  \centering
  \includegraphics{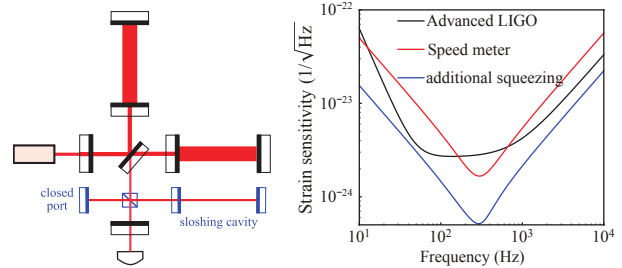}\vspace*{-2mm}
  \caption{(Color online) The sloshing-cavity speed meter (left) and its quantum noise curve (right). With radiation pressure noise cancelled, the quantum noise can be reduced over a broadband with additional constant squeezing.}
   \label{fig:config_Sh_SM}
\end{figure}

Slightly different from the motivations behind frequency-dependent squeezing and readout,
the speed meter originates from the perspective of viewing the GW detector as a quantum measurement device. The speed is proportional to momentum---the conserved dynamical quantity of free test mass. According to quantum measurement theory, measuring conserved quantity is immune to measurement back action noise (in our context the radiation-pressure noise). There are many realizations of speed meter: the Sagnac interferometer\,\cite{Beyersdorf1999, Che2003, Danilishin2004}, different polarizations\,\cite{Wade2012}, and more recently the intra-cavity filtering scheme\,\cite{Wang2014}. Here we consider the one using sloshing filter cavity proposed in ref.\,\cite{Purdue2}, which is shown schematically in Figure\,\ref{fig:config_Sh_SM}. The principle behind goes as follows: the optical field that contains information of the test mass position at earlier moments gets stored in the \linebreak
\vspace*{-3mm}

\noindent sloshing (filter) cavity. It coherently superimposes with (a minus sign) the optical field having current position information, giving rise to speed response. One key feature is the low-frequency part of the noise curve (in Figure\,\ref{fig:config_Sh_SM}) that rolls off as $\Omega^{-1}$ ($\Omega$ being the frequency), in contrast to the position meter with $\Omega^{-2}$.

\subsubsection{Long signal recycling}\label{subsec:LSR}

\begin{figure}[H]
\centering
  \includegraphics{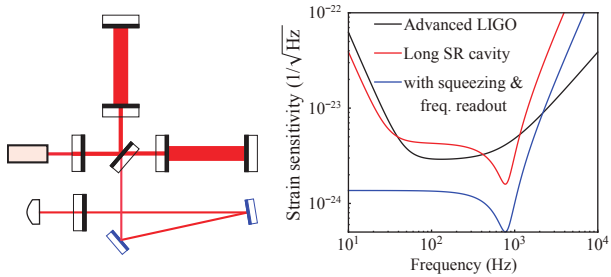}\vspace*{-2mm}
  \caption{(Color online) Schematics for the long signal-recycling cavity scheme (left) and its quantum noise curve in the tuned case (right). General detuned case is more complicated.}
   \label{fig:config_Sh_LSR}
\end{figure}

In the usual case when the beam splitter and the signal recycling mirror are close to each other, the signal recycling cavity is relatively short (order of 10 m) and one can ignore the phase accumulated in this cavity by the audio sidebands.
We can therefore treat the signal-recycling cavity as an effective compound mirror with complex transmissivity and reflectivity, which is the approach applied in ref.\,\cite{BuCh2003}.
A long signal recycling cavity, shown schematically in Figure\,\ref{fig:config_Sh_LSR}, however, allows for
different sidebands picking up different phase shifts. The general scenarios for this scheme is quite complicated. One interesting special case is when the signal-recycling cavity is tuned. The coupled cavity, formed by the signal-recycling cavity and the arm cavity, has two resonant frequencies located symmetrically around the carrier frequency with their
frequency separation determined by the ITM transmissivity. This case has two advantages: firstly, the interferometer has equal response to the upper and lower sideband signals, but at the same time allows for resonant enhancement at high frequencies, shown by the noise curve in Figure\,\ref{fig:config_Sh_LSR}. In contrast to the case with short signal-recycling cavity, detuning the signal-recycling cavity can only resonantly enhance either upper sideband or lower sideband---such an imbalance in the response increases the shot noise by a factor of four in power compared to the balanced case; secondly, there is no optical-spring effect, as the
signal-recycling cavity is tuned, and the test-mass dynamics is therefore not modified. Such a case corresponds to the twin signal-recycling scheme theoretically studied by
Th\"{u}ring et al.\,\cite{Thuring2007} and experimentally demonstrated by Gr\"{a}f et al.\,\cite{Graf2013}, which are motivated by the above mentioned two advantages. One can refer to ref.\,\cite{McClelland1995} and references therein for an overview of different recycling techniques applied in the context of GW detectors.

\subsubsection{Dual-carrier scheme}\label{subsec:DC}

\begin{figure}[H]
\centering
  \includegraphics{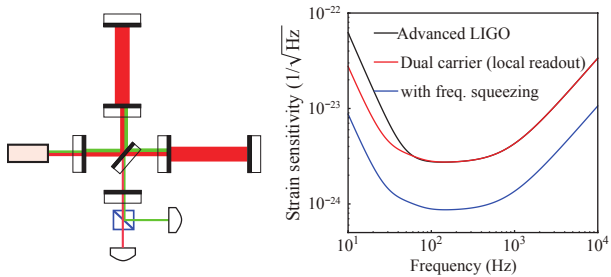}\vspace*{-2mm}
       \caption{(Color online) Schematics for the dual carrier scheme with the auxiliary carrier only resonant inside the power-recycling cavity (anti-resonant with respect to the arm cavity)---the so-called local readout scheme (left) and its quantum noise curve in the tuned case (right).}
   \label{fig:config_Sh_DC}
\end{figure}

The dual-carrier scheme, as shown in Figure\,\ref{fig:config_Sh_DC}, includes an auxiliary
carrier into the interferometer. With its central frequency significantly different from the main one, it
provides us with an additional sensing channel. We can further introduce frequency-dependent squeezing or readout on top of the scheme. The auxiliary carrier can be resonant inside the arm cavity together with the main carrier \cite{Yanbei:DoubleSpring}, or only resonant in the power-recycling cavity which is the so-called local readout scheme\,\cite{Yanbei:Local}. When the signal-recycling cavity is tuned to both carriers, the auxiliary channel will not directly carry information about the GW signal. However, it contains information about the radiation pressure noise of the main carrier by measuring the displacement of the ITMs, therefore allowing for canceling the radiation pressure noise, if we optimally combine the readouts of two carriers, which is shown by the difference between the black curve and red curve at low frequencies shown in Figure\,\ref{fig:config_Sh_DC}. Here we only consider introducing one additional carrier. In principle, one can study general multi-carrier scheme, and an interesting case with multiple paired carriers has been recently studied in ref.\,\cite{Korobko2015}.

\subsection{One comparison of these techniques}

These above mentioned techniques can be combined in various ways to enhance the detector sensitivity and in most
cases, are complimentary to each other. Therefore, there \linebreak
\vspace*{-3mm}

\begin{figure}[H]
\centering
  \includegraphics[scale=0.7]{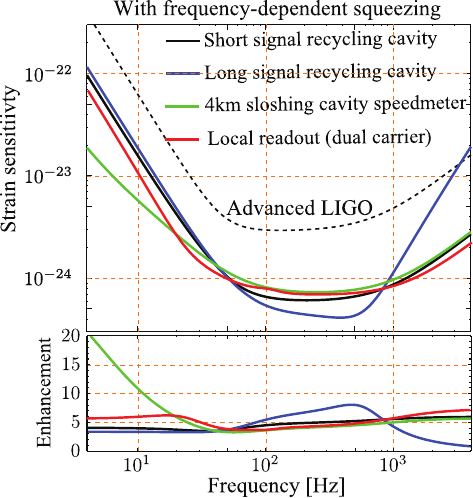}
       \caption{(Color online) Optimization results for the quantum noise curves of different schemes with frequency dependent squeezing (using a single filter cavity of 100 m and 30 ppm round-trip loss). The enhancement factor is defined as the ratio between the quantum noise spectrum (in amplitude) of aLIGO (RSE mode) and that of the listed schemes. The test mass is assumed to be 150 kg, maximal arm cavity power to be 3 MW (the cavity power is an optimization parameter), and 10 dB squeezing in these schemes, in order to get around a factor of five enhancement over a broad frequency band using achievable parameters. }
   \label{fig:comparison_FDS}
\end{figure}

\begin{figure}[H]
\centering
  \includegraphics[scale=0.7]{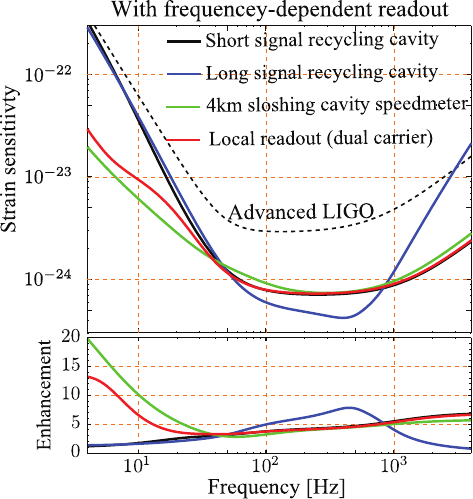}
       \caption{(Color online) Optimization results for the quantum noise of different schemes with frequency-dependent readout. Other specifications are the same as Figure\,\ref{fig:comparison_FDS}. Given the 100 m scale filter cavity and 30 ppm loss, the low-frequency performance of frequency-dependent readout becomes comparable to that of frequency-dependent squeezing, however it can be better if the cavity length becomes longer or the loss is lower. }
   \label{fig:comparison_FDR}
\end{figure}

\noindent is no single figure of merit that allows us to compare them on an equal footing. In ref.\,\cite{haixing}, one attempt to compare them is made by assuming a cost function that tries to maximize the broadband enhancement over aLIGO in the RSE mode. In Figure\,\ref{fig:comparison_FDS}, we show the optimized quantum noise for different schemes with frequency-dependent squeezing included in all of them, and the filter cavity is a single 100 m scale cavity (one for each carrier in the dual-carrier scheme) with realistic round-trip loss of 30 ppm. In Figure\,\ref{fig:comparison_FDR}, we show the noise curves of different schemes with \linebreak
\vspace*{-3mm}

\noindent frequency-dependent readout and squeezing injection (constant angle). The specification for the filter cavity is the same as the one in frequency-dependent squeezing. As we can see, the short signal-recycling cavity interferometer (aLIGO configuration) with frequency-dependent squeezing is one promising candidate for near-future upgrade in terms of complexity. Further low-frequency enhancement can be achieved by using speed meter or local-readout scheme which however involves some additional complexity.

\vspace*{2mm}
\Acknowledgements{\bahao The authors  thank the Kavli Institute for Theoretical Physics, China for funding the Next Detectors for Gravitational Astronomy Program, and for their hospitality. The authors of sect. 1 particularly thank Prof. WU YueLiang for many useful and interesting discussions. The author of sect. 2 gratefully acknowledges the support of the United States National Science Foundation for the construction and operation of the LIGO Laboratory and the Science and Technology Facilities Council of the United Kingdom, the Max-Planck-Society, and the State of Niedersachsen/Germany for support of the construction and operation of the GEO600 detector. The author also gratefully acknowledges the support of the research by these agencies and by the Australian
Research Council, the Council of Scientific and Industrial Research of India, the Istituto Nazionale di Fisica Nucleare
of Italy, the Spanish Ministerio de Econom\'ia y Competitividad, the Conselleria d'Economia Hisenda i Innovaci\'o of the
Govern de les Illes Balears, the Royal Society, the Scottish Funding Council, the Scottish Universities Physics
Alliance, the National Aeronautics and Space Administration, the Carnegie Trust, the Leverhulme Trust, the David and
Lucile Packard Foundation, the Research Corporation, and the Alfred P. Sloan Foundation. S.H. acknowledges the support
from the European Research Council (ERC-2012-StG: 307245). This paper has been assigned LIGO document no.\
LIGO-P1300081.  BLAIR C.  would like to particularly thank the staff at LIGO Livingston particularly Valery Frolov, Ryan DeRosa, Joseph Betzwieser, Adam Mallavey and Brian O'Reilly who made the investigations of PI and control at LIGO Livingston possible and the oversight from Peter Fritschel, Slawomir Gras and Aidan Brooks. BLAIR C. thanks colleagues at ACIGA for help along the way. This work was supported by the LSC LIGO visitor program, the Australian Department of Education and Australian Research Council. This work was also supported by
Australian Research Council (Grant Nos. DP120100898 and DP120104676). LIGO was constructed by the California Institute of Technology and Massachusetts Institute of Technology with funding from the National Science Foundation, and operates under cooperative agreement PHY-0757058. Advanced LIGO was built under award PHY-0823459. H. M. is supported by the Marie-Curie Fellowship.}

\end{multicols}

\end{document}